\documentclass[12pt]{iopart}
\begin{document}
\jl{4}
\title[QED RC to the two identical fermion scattering]
{The QED lowest-order radiative corrections to the two polarized
identical
fermion scattering}
\author {N M Shumeiko and J G Suarez}
\address{National Center of Particle and
 High Energy Physics, Bogdanovich Str 153, 220040 Minsk, Belarus}
\begin {abstract}
The explicit exact expressions of QED lowest-order radiative
corrections 
to the two polarized identical fermion scattering are
 presented in covariant form. Polarization
effects 
are treated in detail.    
 The infrared divergence from the
real photon emission is extracted by covariant approach.
 Some numerical results for M{\o}ller
scattering are given.
 
\end {abstract}
\pacs{12.15.Lk, 13.60.-r, 13.88.+e}
\submitted
\maketitle

\input{epsf}

\section{Introduction}
The two identical fermion scattering has been
investigated since the arising of {\it Quantum
Electrodinamics}. In the early 1930s M{\o}ller
\cite{Moller} calculated the unpolarized cross section of
this process at the Born level. 
The lowest-order radiative corrections (RC)
to the two identical fermion scattering were first calculated by
Redhead \cite{Redhead}, Polovin \cite{Pol1, Pol2} and Tsai
\cite{Tsai} for the unpolarized cross section. DeRaad \cite{Deraad1,
Deraad}, using
the approaches proposed by Mo and Tsai \cite{MoT} and
Gastmans and collaborators \cite{Cal}, carried out a calculation of RC  
taking into account particle polarization.
However these results  
 are set-up dependent (the unphysical parameter $\Delta E$ 
appears in the results) and the contribution of ``hard
photons'' has been neglected. In \cite{Denner} an exhaustive analysis
of 
RC is presented but the results are also set-up dependent.
 In \cite{Jadach} Monte Carlo method was used for calculation of RC at
SLAC set-up. Some separate contributions were widely discussed in 
\cite{Kuraev, Cuypers, Czarnecki}.

Thus, we can conclude that 
 the results of calculations of RC to the two polarized identical
fermion
scattering have been presented in non-covariant form, they were oriented
mainly, to the study of scattering in the center of mass system and an
exhaustive analysis of polarization effects has been not performed.  
 The aim of this paper is to present a complete
QED lowest-order calculation of RC to cross section 
and other observables in the process of the two  
polarized identical fermion scattering in covariant form. We propose
formulae,
which can be used both 
 for numerical analysis of fixed target experiments
and  collider ones. A detailed analysis of polarization effects
 is presented. Since we consider the scattering of any two identical
fermions it is desirable to have exact formulae, without approximations
such as ultrarelativistic (URA) or leading log (LLA). One of the main
applications of proposed results is the study
of RC to polarized M{\o}ller scattering \cite{Moller} and, in particular,
to M{\o}ller
polarimeter.
    
 It is known that two identical fermion scattering
is characterized
by the presence of t- and u- channel diagrams 
 and their interference.
That fact makes the calculation of cross section more
complicated than for two non-identical fermion scattering
(e.g. $\mu^{-}e^{-}$), where only the contribution of one of
these channels have to
 be considered and interference vanishes. 
The Feynman graphs, which contribute to the Born cross section of 
the considered process
 are shown on figure 1.
The full set of Feynman graphs which is necessary to
calculate the
QED lowest-order RC  is
presented on figure 2.

The observed cross section is given by
\begin{equation}
\begin{array}{ll}
\displaystyle
 {\sigma_{obs}} \sim& 
\displaystyle
  \vert{M_{a}+M_{b}}\vert^{2}+2Re
 (M_{a}+M_{b})\sum_{i,i'=1}^{5}(M_{i}+M_{i'})^{+}+
\\[0.5cm]
&\displaystyle
+ {\sum_{i,i'=6}^{9}\vert(M_{i}+ M_{i'})}\vert^{2}.
\end{array}
\label{fff}
\end{equation}
Here $M_{a(b)}$ are the matrix elements of
Born (or one-photon exchange)
contribution.
$M_{i(i')}$ are the matrix elements of radiative processes. 
In order to calculate exactly the QED lowest-order RC to the 
 two polarized identical fermion scattering, the method offered
in ref.\cite{BS}
is used (for more
details see also
\cite{Prep1}-\cite{Prep4}). 

This paper is organized as follows.
In section 2 the kinematics for elastic
 process (Born and
virtual photon(V) contributions) and
 inelastic one (real photon(R) contribution) is
presented.
 Section 3 is devoted to the presentation of explicit
formulae for V- and R-contributions.  
We conclude with an illustration and     
 short
 discussion of obtained results in section 5.

%\newpage
\begin{figure}[h]
\begin{center}
%\vspace{-1cm}
%\hspace{-1cm}
\begin{tabular}{cc}
\begin{picture}(100,100)
\put(30,61){\line(2,-1){20.}}
\put(50,51){\line(2,1){20.}}
\put(50,51){\circle*{2.}}
\put(30,8){\line(2,1){20.}}
\put(50,18){\line(2,-1){20.}}
\put(50,18){\circle*{2.}}
\multiput(50,20)(0,8){4}{\oval(4.0,4.0)[r]}
\multiput(50,24)(0,8){4}{\oval(4.0,4.0)[l]}
\put(40,35){\makebox(0,0){\footnotesize $\gamma $}}
%\put(31,48){\makebox(0,0){\footnotesize $\it e^{-} $}}
%\put(72,48){\makebox(0,0){\footnotesize $\it e^{-} $}}
%\put(31,22){\makebox(0,0){\footnotesize $\it e^{-} $}}
%\put(72,22){\makebox(0,0){\footnotesize $\it e^{-} $}}
\put(20,60){\makebox(0,0){\footnotesize $k_{1}$}}
\put(80,60){\makebox(0,0){\footnotesize $k_{2}$}}
\put(20,10){\makebox(0,0){\footnotesize $p_{1}$}}
\put(80,10){\makebox(0,0){\footnotesize $p_{2}$}}
\put(50,-2){\makebox(0,0){\small (a) }}
\end{picture}
&
\begin{picture}(100,100)
\put(30,61){\line(2,-1){20.}}
\put(50,51){\line(4,-3){40.}}
\put(50,51){\circle*{2.}}
\put(30,8){\line(2,1){20.}}
\put(50,18){\line(4,3){40.}}
\put(50,18){\circle*{2.}}
\multiput(50,20)(0,8){4}{\oval(4.0,4.0)[r]}
\multiput(50,24)(0,8){4}{\oval(4.0,4.0)[l]}
\put(40,35){\makebox(0,0){\footnotesize $\gamma $}}
\put(20,60){\makebox(0,0){\footnotesize $k_{1}$}}
\put(85,55){\makebox(0,0){\footnotesize $k_{2}$}}
\put(20,10){\makebox(0,0){\footnotesize $p_{1}$}}
\put(85,15){\makebox(0,0){\footnotesize $p_{2}$}}
\put(56,-2){\makebox(0,0){\small (b)}}
\end{picture}
\end{tabular}
\end {center}
\caption{Feynman diagrams contributing to the Born cross
section. t-channel (a) and u-channel (b).}
\end{figure}
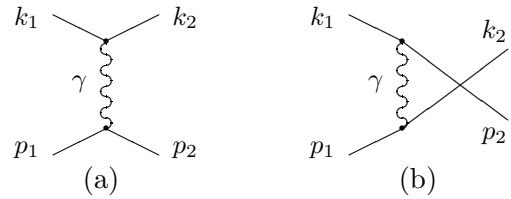
%\begin{figure}[h]
%\vspace{-1cm}
%\hspace{-1cm}
\begin{center}
\begin{tabular}{ccc}
\begin{picture}(100,100)
\put(30,60){\line(2,-1){20.}}
\put(50,50){\line(2,1){20.}}
\put(30,10){\line(2,1){20.}}
\put(50,20){\line(2,-1){20.}}
\put(50,50){\circle*{2.}}
\put(50,20){\circle*{2.}}
\multiput(50,48)(0,-8){1}{\oval(4.0,4.0)[r]}
\multiput(50,44)(0,-8){1}{\oval(4.0,4.0)[l]}
\put(50,41){\circle*{2.}}
\put(50,28){\circle*{2.}}
\multiput(50,26)(0,8){1}{\oval(4.0,4.0)[r]}
\multiput(50,22)(0,8){1}{\oval(4.0,4.0)[l]}
\put(50,35){\circle{12.}}
%\put(40,44){\makebox(0,0){\footnotesize $\gamma $}}
%\put(40,26){\makebox(0,0){\footnotesize $\gamma $}}
\put(31,48){\makebox(0,0){\footnotesize $k_{1} $}}
\put(82,48){\makebox(0,0){\footnotesize $k_{2}(p_{2}) $}}
\put(31,22){\makebox(0,0){\footnotesize $p_{1} $}}
\put(82,22){\makebox(0,0){\footnotesize $p_{2}(k_{2}) $}}
\put(50,-3){\makebox(0,0){\footnotesize $ (1,1') $}}
\end{picture}
&
\begin{picture}(100,100)
\put(30,61){\line(2,-1){20.}}
\put(50,51){\line(2,1){20.}}
\put(50,51){\circle*{2.}}
\put(30,8){\line(2,1){20.}}
\put(50,18){\line(2,-1){20.}}
\put(50,18){\circle*{2.}}
\multiput(50,20)(0,8){4}{\oval(4.0,4.0)[r]}
\multiput(50,24)(0,8){4}{\oval(4.0,4.0)[l]}
\put(50,65.5){\oval(4.0,4.0)[t]}
\put(39,56.5){\circle*{2.}}
\put(61,56.5){\circle*{2.}}
\put(45,59.5){\oval(6.0,6.0)[lt]}
\multiput(39,59.5)(6,6){2}{\oval(6.0,6.0)[br]}
\put(55,59.5){\oval(6.0,6.0)[rt]}
\multiput(55,65.5)(6.0,-6.0){2}{\oval(6.0,6.0)[lb]}
%\put(31,48){\makebox(0,0){\footnotesize $k_{1} $}}
%\put(82,48){\makebox(0,0){\footnotesize $k_{2}(p_{2}) $}}
%\put(31,22){\makebox(0,0){\footnotesize $p_{1} $}}
%\put(82,22){\makebox(0,0){\footnotesize $p_{2}(k_{2}) $}}
\put(50,-3){\makebox(0,0){\footnotesize $ (2,2') $}}
\end{picture}
&
\begin{picture}(100,100)
\put(30,61){\line(2,-1){20.}}
\put(50,51){\line(2,1){20.}}
\put(50,51){\circle*{2.}}
\put(30,9){\line(2,1){20.}}
\put(50,19){\line(2,-1){20.}}
\put(50,19){\circle*{2.}}
\multiput(50,21)(0,8){4}{\oval(4.0,4.0)[r]}
\multiput(50,25)(0,8){4}{\oval(4.0,4.0)[l]}
\put(50,4.5){\oval(4.0,4.0)[b]}
\put(39,13.5){\circle*{2.}}
\put(61,13.5){\circle*{2.}}
\put(45,10.75){\oval(6.0,6.0)[lb]}
\multiput(39,10.5)(6.0,-6.0){2}{\oval(6.0,6.0)[tr]}
\put(55,10.75){\oval(6.0,6.0)[rb]}
\multiput(55,4.5)(6.0,6.0){2}{\oval(6.0,6.0)[lt]}
%\put(31,48){\makebox(0,0){\footnotesize $k_{1} $}}
%\put(82,48){\makebox(0,0){\footnotesize $k_{2}(p_{2}) $}}
%\put(31,22){\makebox(0,0){\footnotesize $p_{1} $}}
%\put(82,22){\makebox(0,0){\footnotesize $p_{2}(k_{2}) $}}
\put(50,-4){\makebox(0,0){\footnotesize $ (3,3') $}}
\end{picture}
\end{tabular}
\end{center}
%\end{figure}
%\begin{figure}[h]
%\vspace{-1cm}
%\hspace{-1cm}
\begin{center}
\begin{tabular}{cc}
\begin{picture}(100,100)
\put(30,60){\line(2,-3){10.}}
\put(40,45){\line(2,0){20.}}
\put(60,45){\line(2,3){10.}}
\put(30,10){\line(2,3){10.}}
\put(40,25){\line(2,0){20.}}
\put(60,25){\line(2,-3){10.}}
\multiput(42,25)(4,4){5}{\oval(4.0,4.0)[lt]}
\multiput(42,29)(4,4){5}{\oval(4.0,4.0)[br]}
\multiput(42,45)(4,-4){5}{\oval(4.0,4.0)[lb]}
\multiput(42,41)(4,-4){5}{\oval(4.0,4.0)[tr]}
\put(38,35){\makebox(0,0){\footnotesize $\gamma$}}
\put(62,35){\makebox(0,0){\footnotesize $\gamma$}}
%\put(25,52){\makebox(0,0){\footnotesize $k_{1} $}}
%\put(90,52){\makebox(0,0){\footnotesize $k_{2}(p_{2}) $}}
%\put(25,24){\makebox(0,0){\footnotesize $p_{1} $}}
%\put(90,24){\makebox(0,0){\footnotesize $p_{2}(k_{2}) $}}
\put(50,-3){\makebox(0,0){\footnotesize $ (4,4') $}}
\end{picture}
&
\begin{picture}(100,100)
\put(30,60){\line(2,-3){10.}}
\put(40,45){\line(2,0){20.}}
\put(60,45){\line(2,3){10.}}
\put(30,10){\line(2,3){10.}}
\put(40,25){\line(2,0){20.}}
\put(60,25){\line(2,-3){10.}}
\multiput(40,27)(0,8){3}{\oval(4.0,4.0)[l]}
\multiput(40,31)(0,8){2}{\oval(4.0,4.0)[r]}
\multiput(60,27)(0,8){3}{\oval(4.0,4.0)[r]}
\multiput(60,31)(0,8){2}{\oval(4.0,4.0)[l]}
\put(33,35){\makebox(0,0){\footnotesize $\gamma$}}
\put(67,35){\makebox(0,0){\footnotesize $\gamma$}}
%\put(25,52){\makebox(0,0){\footnotesize $k_{1} $}}
%\put(90,52){\makebox(0,0){\footnotesize $k_{2}(p_{2}) $}}
%\put(25,24){\makebox(0,0){\footnotesize $p_{1} $}}
%\put(90,24){\makebox(0,0){\footnotesize $p_{2}(k_{2}) $}}
\put(50,-3){\makebox(0,0){\footnotesize $ (5,5') $}}
\end{picture}
\end{tabular}
\end{center}
%\end{figure}
%\\[0.2cm]
\begin{figure}[h]
\begin{center}
%\vspace{-1cm}
%\hspace{-1cm}
\begin{tabular}{cccc}
\begin{picture}(100,100)
\multiput(40,57)(0,8){3}{\oval(4.0,4.0)[r]}
\multiput(40,61)(0,8){3}{\oval(4.0,4.0)[l]}
\put(41,56){\circle*{2.}}
\put(30,61){\line(2,-1){20.}}
\put(50,51){\line(2,1){20.}}
\put(50,51){\circle*{2.}}
\multiput(50,20)(0,8){4}{\oval(4.0,4.0)[r]}
\multiput(50,24)(0,8){4}{\oval(4.0,4.0)[l]}
\put(30,8){\line(2,1){20.}}
\put(50,18){\line(2,-1){20.}}
\put(50,18){\circle*{2.}}
%\put(49,20){\line(0,2){30.}}
%\put(51,20){\line(0,2){30.}}
\put(31,48){\makebox(0,0){\footnotesize $k_{1} $}}
\put(82,48){\makebox(0,0){\footnotesize $k_{2}(p_{2}) $}}
\put(31,22){\makebox(0,0){\footnotesize $p_{1} $}}
\put(82,22){\makebox(0,0){\footnotesize $p_{2}(k_{2}) $}}
\put(47,70){\makebox(0,0){\footnotesize $k$}}
\put(50,-3){\makebox(0,0){\footnotesize $ (6,6') $}}
\end{picture}
&
\begin{picture}(100,100)
\multiput(60.5,57.5)(0,8){3}{\oval(4.0,4.0)[l]}
\multiput(60.5,61.5)(0,8){3}{\oval(4.0,4.0)[r]}
\put(61,56){\circle*{2.}}
\put(50,51){\circle*{2.}}
\put(30,61){\line(2,-1){20.}}
\put(50,51){\line(2,1){20.}}
\multiput(50,20)(0,8){4}{\oval(4.0,4.0)[r]}
\multiput(50,24)(0,8){4}{\oval(4.0,4.0)[l]}
\put(30,8){\line(2,1){20.}}
\put(50,18){\line(2,-1){20.}}
\put(50,18){\circle*{2.}}
%\put(31,48){\makebox(0,0){\footnotesize $k_{1} $}}
%\put(82,48){\makebox(0,0){\footnotesize $k_{2}(p_{2}) $}}
%\put(31,22){\makebox(0,0){\footnotesize $p_{1} $}}
%\put(82,22){\makebox(0,0){\footnotesize $p_{2}(k_{2}) $}}
\put(50,-3){\makebox(0,0){\footnotesize $ (7,7') $}}
\end{picture}
&
\begin{picture}(100,100)
\multiput(40,15.5)(0,8){3}{\oval(4.0,4.0)[l]}
\multiput(40,19.5)(0,8){3}{\oval(4.0,4.0)[r]}
\put(41,14){\circle*{2.}}
\put(30,61){\line(2,-1){20.}}
\put(50,51){\line(2,1){20.}}
\put(50,51){\circle*{2.}}
\multiput(50,20)(0,8){4}{\oval(4.0,4.0)[r]}
\multiput(50,24)(0,8){4}{\oval(4.0,4.0)[l]}
\put(30,8){\line(2,1){20.}}
\put(50,18){\line(2,-1){20.}}
\put(50,18){\circle*{2.}}
%\put(49,20){\line(0,2){30.}}
%\put(51,20){\line(0,2){30.}}
%\put(28,48){\makebox(0,0){\footnotesize $k_{1} $}}
%\put(82,48){\makebox(0,0){\footnotesize $k_{2}(p_{2}) $}}
%\put(24,22){\makebox(0,0){\footnotesize $p_{1} $}}
%\put(82,22){\makebox(0,0){\footnotesize $p_{2}(k_{2}) $}}
\put(50,-3){\makebox(0,0){\footnotesize $ (8,8') $}}
\end{picture}
&
\begin{picture}(100,100)
\multiput(60,15.5)(0,8){3}{\oval(4.0,4.0)[r]}
\multiput(60,19.5)(0,8){3}{\oval(4.0,4.0)[l]}
\put(60,14){\circle*{2.}}
\put(30,61){\line(2,-1){20.}}
\put(50,51){\line(2,1){20.}}
\put(50,51){\circle*{2.}}
\multiput(50,20)(0,8){4}{\oval(4.0,4.0)[r]}
\multiput(50,24)(0,8){4}{\oval(4.0,4.0)[l]}
\put(30,8){\line(2,1){20.}}
\put(50,18){\line(2,-1){20.}}
\put(50,18){\circle*{2.}}
%\put(49,20){\line(0,2){30.}}
%\put(51,20){\line(0,2){30.}}
%\put(31,48){\makebox(0,0){\footnotesize $k_{1} $}}
%\put(86,48){\makebox(0,0){\footnotesize $k_{2}(p_{2}) $}}
%\put(31,22){\makebox(0,0){\footnotesize $p_{1} $}}
%\put(90,22){\makebox(0,0){\footnotesize $p_{2}(k_{2}) $}}
\put(50,-3){\makebox(0,0){\footnotesize $ (9,9') $}}
\end{picture}
\end{tabular}
\end{center}
\caption{
Feynman diagrams contributing to the QED lowest-order
radiative corrections  
to the two identical fermion scattering.}
\end{figure}
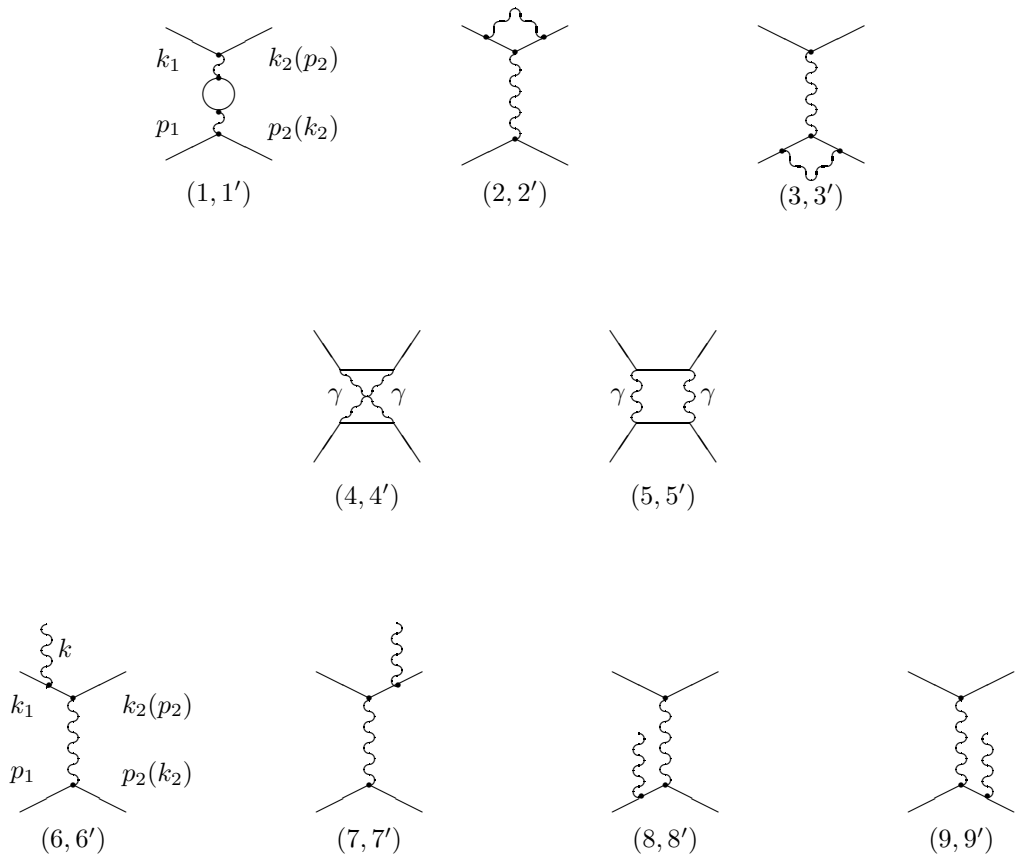
\newpage
\section{Kinematics and phase space}
We consider both the elastic scattering 
\begin{equation}
f_{1}(k_{1},\xi)+f_{2}(p_{1},\eta)\to
f_{3}(k_{2})+f_{4}(p_{2})
\label {ep}
\end{equation}
and the bremsstrahlung process
\begin{equation}
f_{1}(k_{1},\xi)+f_{2}(p_{1},\eta)\to
f_{3}(k_{2})+f_{4}(p_{2})
 +\gamma(k)
\label {brems}
\end{equation}
where $k_{1}$,$p_{1}$($k_{2}$,$p_{2}$) are four momenta of 
initial(final) fermions and $k$ is the four momenta of a real photon.
$\xi$, $\eta$ are the beam and target polarization vectors.

\noindent In the case of 
longitudinal (see \cite{Aku}) polarization
\begin{equation}
\begin{array}{ll}
\displaystyle
\xi=&
\displaystyle
\xi_{l}=\frac{1}{\sqrt{\lambda_{s}}}\biggr(\frac{S}{m}k_{1}
-2mp_{1}\biggl)
\\[0.5cm]
&\displaystyle
\hspace{-1.0cm}
\eta=\eta_{l}=\frac{1}{\sqrt{\lambda_{s}}}\biggr(2mk_{1}
-\frac{S}{m}p_{1}\biggl)
\end{array}
\label {par}
\end{equation}
%\end{center}
where $S=2p_{1}.k_{1}$,\; $\lambda_{s}=S^{2}-4m^{4}$,
and\;\; $m$\; is the fermion mass.
If the beam and target are polarized
transversally 
\begin{equation}
\begin{array}{ll}
\displaystyle 
\xi=&
\displaystyle 
\xi_{t}=\frac{1}{{\cal K} \sqrt{ \lambda_{s}}}\biggr((-SX_{o}+2m^{2}Q^{2}_{m})k_{1}
+\lambda_{s}k_{2}-(SQ^{2}+2m^{2}S^{o}_{x})p_{1}
\biggl)
\\[0.5cm]
&\displaystyle
\hspace{-1.0cm}
\eta=\eta_{t}=\xi_{t}.
\end{array}
\label {par1}
\end{equation}
Here
$Q^{2}=-(k_{1}-k_{2})^{2}$,\;\;$Q^{2}_{m}=Q^{2}+2m^{2}$,\;\;$X_{o}=2p_{1}.k_{2}$,\;\;
$S^{o}_{x}=S-X_{o}$\;\;and\;\; 
\newline
${\cal K}=(Q^{2}(SX_{o}-2m^{2}Q^{2}_{m}))^{\frac{1}{2}}$.
Finally, if the beam and target are normal\footnote{ normal it 
means perpendicular to the scattering plane.} polarized
\begin{equation}
\begin{array}{ll}
\displaystyle 
\xi^{\mu}=&
\displaystyle 
\xi_{\bot}^{\mu}=\frac{1}{\cal K}\epsilon_{\mu \lambda \gamma
\theta}k_{1}^{\lambda}k_{2}^{\gamma}p_{1}^{\theta} 
\\[0.5cm]
&\displaystyle
\hspace{-1.0cm}
\eta^{\beta}=\eta_{\bot}^{\beta}=\frac{1}{\cal K}\epsilon_{\beta \lambda \gamma
\theta}k_{1}^{\lambda}k_{2}^{\gamma}p_{1}^{\theta} 
\end{array}
\label {par2}
\end{equation}
\begin{equation}
\eta^{2}=\eta_{\bot}^{2}=\xi_{\bot}^{2}=-1
\nonumber
\end{equation}
where $\epsilon_{\mu \lambda \gamma \theta}$ is the Levi-Civita's
tensor.

\noindent The cross section of (\ref {brems}) depends on five
kinematical variables. We use the independent set:
\begin{center}
$S$,\;\; 
$y=\displaystyle\frac{S^{o}_{x}}{S}$,\;\;
$v=2p_{2}.k$,\;\;
$t=-(p_{2}-p_{1}-k)^{2}$,\;\; 
$z_{2}=2k_{2}.k$. 
\end{center}

\noindent Using the parametrizations (\ref {par}), (\ref {par1}) and (\ref
{par2}), a general expression of
the Born cross section of the two polarized identical fermion scattering
can be presented in the form\footnote{Here and below
$\sigma=\frac{d\sigma}{dy}.$}:
\begin{equation}
\begin{array}{ll}
\displaystyle
\sigma_{o}=&
\displaystyle
\frac{4\pi\alpha^{2}S}{\lambda_{s}} \biggr[
\biggr(\frac{U_{1}}{Q^{4}}+\frac{U_{2}}{Q_{1}^{4}}+
\frac{U_{3}}{Q^{2}Q_{1}^{2}} \biggl)+
\\[0.5cm]
&\displaystyle
+ \sum_{B} \sum_{T}
P_{B}P_{T} 
\biggr(\frac{P_{1BT}}{Q^{4}}+\frac{P_{2BT}}{Q_{1}^{4}}+
\frac{P_{3BT}}{Q^{2}Q^{2}_{1}} \biggl) \biggl].
\label{born}
\end{array}
\end{equation}
In (\ref {born}) 
indexes $B$(beam) and $T$(target) mean $B$=$T$=$l$(long.),
$t$(transv.), $\bot$(normal). The quantities $U_{i}$ and $P_{iBT}$
are given in explicit form in {\bf appendix A}, 
$Q^{2}_{1}=X_{o}-2m^{2}$ and  
 $P_{B(T)}$ are the polarization degree of beam(target) fermions. 

\section{Exact formulae for the lowest-order radiative corrections}
 The cross section of analysed process 
 in order $\alpha^{3}$ is given by 
\begin{eqnarray}
%\begin{array}{ll}
%\displaystyle
  \sigma=
%\displaystyle
   \sum_{k}(\sum_{j}
   \sigma_{k}^{u,j} + 
P_{B}P_{T}\sum_{j}
   \sigma_{k}^{p,j}) +
%   \\[0.5cm]
%&\displaystyle
    \sum_{h}(\sum_{n}
   \sigma_{h}^{u,n} + P_{B}P_{T}\sum_{n}
   \sigma_{h}^{p,n})).
\label {csm}
%\end{array}
\end{eqnarray}
\noindent Here {\it j} runs over contributions of vacuum
polarization,
vertex diagrams and bremsstrahlung.
The index {\it k} points out sum over the contributions of t-($l$),
u-($e$) channel diagrams and their interference($i$).
 The index {\it n} denotes sum over
anomalous magnetic moment and two-photon exchange. For these last
corrections the contribution of the interference of t- and u- channel
diagrams has been divided into two parts. For this reason we  have separated these
corrections from the other ones and use the index {\it h} instead of {\it
k} (see bellow formulas (\ref{amm}) and (\ref{two})). The index
{\it u}({\it p})
is used to separate the contribution of unpolarized(polarized) parts.
\subsection{V-contribution}
We call ``factorized RC'' the sum of all $\delta$ which are
looked
like that
$\sigma=\delta \sigma_{o}$. 
 It is very convenient,
for
numerical analysis, to collect together all contributions with factorized
corrections  
\begin{equation}
  \sigma^{u(p)}_{fact}=
%\displaystyle
 \frac{\alpha}{\pi}\sum_{k} \sum_{f}\delta_{f}^{k}\sigma_{o}^{u(p),k}.
\label {fact}
\end{equation}
In (\ref{fact}) 
$\sum_{k} \sigma_{o}^{u(p),k}$ are the unpolarized(polarized) parts of
 the Born cross section (\ref{born}).
The index ${\it f}$ 
means:
$f$=$vp$, $vert$, $L$, $\lambda$, $s$ ,$K$. It is evident from
(\ref{fact}), that corrections $\delta^{k}_{f}$
  do not depend on polarization.    

 The quantity $\delta_{vp}$ is the well known correction due to
vacuum
polarization
(see figure 2(1,1$'$))
\begin{equation}
\delta_{vp}^{l}=\sum_{i=e,\mu,\tau,...}q_{i}^{2}\biggr[\frac{2}{3}
 \biggr(Q^{2}+2m_{i}^{2}\biggl)L_{m_{i}}-\frac{10}{9}+\frac{8m_{i}^{2}}
{3Q^{2}}\biggr(1-2m_{i}^{2}L_{m_{i}}\biggl)\biggl]
\label{vp}
\end{equation}
where
\begin{eqnarray}
\fl
\hspace{2.50cm} 
L_{m}=\frac{1}{\sqrt{\lambda_{m}}}\ln\frac{\sqrt{\lambda_{m}}+Q^{2}}
{\sqrt{\lambda_{m}}-Q^{2}}\;\;\;\;\;\qquad&
\lambda_{m}=Q^{2}(Q^{2}+4m^{2})
\nonumber
\end{eqnarray}
and
\begin{equation}
\delta_{vp}^{i}=\delta_{vp}^{l}+\delta_{vp}^{e}
\end{equation}
where\;\; $\delta_{vp}^{e}$$=$$\delta_{vp}^{l}$($Q^{2} \to Q^{2}_{1}$).
 In  (\ref {vp}) $q_{i}$  
is the charge of leptons or quarks.
The quantity $\delta_{vert}$ is the correction due to convergent part
of vertex diagrams (see figure 2(2,2$'$; 3,3$'$))
\begin{equation}
\delta_{vert}^{l}=2\biggr(\frac{3}{2}Q^{2}+4m^{2}\biggl)L_{m}-4
\end{equation}
\begin{equation}
\delta_{vert}^{i}=\delta_{vert}^{l}+\delta_{vert}^{e}
\end{equation}
where\;\; $\delta_{vert}^{e}$$=$$\delta_{vert}^{l}$($Q^{2} \to
Q^{2}_{1}$).
 The correction $\delta_{L}$, obtained from the sum of finite terms
derived
from infrared divergent part of vertex and two-photon exchange
diagrams, is: 
\begin{equation}
\delta_{L}^{l}=2[L(p_{1},k_{1})-L(k_{1},k_{2})-L(p_{1},k_{2})]
\end{equation}
\begin{equation}
\delta_{L}^{l}=\delta_{L}^{e}=\delta_{L}^{i}.
\end{equation}
The explicit expressions of $L(p_{1},k_{1})$, $L(k_{1},k_{2})$
and $L(p_{1},k_{2})$ are given in \cite{BS} (see (66)). 
The correction $\delta_{\lambda}$, coming from the sum
of infrared
divergent
parts of
V- and R-contributions, is:
\begin{equation}
\delta_{\lambda}^{l}=J(Q^{2},0)\ln\frac{v_{max}}{2m^{2}}
%\label {dlamb}
\end{equation}
\begin{equation}
\delta_{\lambda}^{l}=\delta_{\lambda}^{e}=\delta_{\lambda}^{i}.
\end{equation}
Here
$$
\hspace*{-3.0cm}
J(Q^{2},0)=2[2(Q^{2}_{m}L_{m}-1)+X_{o}L_{X_{o}}-SL_{S}]
$$
$$
\hspace*{-6.50cm}
L_{S}=\frac{1}{\sqrt{\lambda_{s}}}\ln\frac{S+\sqrt{\lambda_{s}}}
{S-\sqrt{\lambda_{s}}} 
$$
$$
\hspace*{-2.5cm}
 L_{X_{o}}=L_{S}(S \to -X_{o}),\;\;\; 
\lambda_{x_{o}}=\lambda_{s}(S \to -X_{o})
$$
$$
\hspace*{-.60cm}
v_{max}=\frac{2Q^{2}d_{s}(Q^{2}_{max}-Q^{2})}{Q^{2}d_{s}+\sqrt{\lambda_{m} 
\lambda_{s}}},\;Q^{2}_{max}=\frac{\lambda_{s}}{d_{s}}, \;d_{s}=S+2m^{2}.
$$
The quantity $\delta_{s}$ is the finite part of the "soft-photon``
contribution:
\begin{equation}
\delta^{l}_{s}=\delta^{e}_{s}=\delta^{i}_{s}.
\end{equation}
The explicit expression for $\delta_{s}$ was given in \cite{BS} (see 
(56)).
 The correction $\delta_{K}$  
 follows
from diagrams of two-photon exchange. Its explicit form  
 was presented first in \cite{Kah} (see (A45)).
      
\noindent The contribution of the anomalous magnetic moment (see
 figure 2(2,2$'$; 3,3$'$)) can be expressed as 
\begin{equation}
\begin{array}{ll}
\displaystyle
\sigma_{amm}=&
\displaystyle
 \frac{8\pi\alpha^{2}m^{2}S}{\lambda_{s}}\frac{\alpha}{\pi}\biggr[
 \biggr(\frac{A_{1}}{Q^{2}}+\frac{A_{2}}{Q^{2}Q^{2}_{1}}\biggl)L_{m}+
 \biggr(\frac{A_{3}}{Q_{1}^{4}}+\frac{A_{4}}{Q^{2}Q^{2}_{1}}\biggl)L_{mex}+
   \\[0.5cm]
&\displaystyle
+\sum_{B} \sum_{T}P_{B}P_{T}
\biggr(\biggr(\frac{A_{5BT}}{Q^{2}}+\frac{A_{6BT}}{Q^{2}_{1}}\biggl)L_{m}+
   \\[0.5cm]
&\displaystyle
+\biggr(\frac{A_{7BT}}{Q_{1}^{4}}+\frac{A_{8BT}}{Q^{2}Q^{2}_{1}}\biggl)L_{mex}\biggl)\biggl].
\end{array}
\label{amm}
\end{equation}
Here
$\lambda_{mex}=\lambda_{m}(Q^{2} \to Q^{2}_{1})$,
   $L_{mex}=L_{m}(Q^{2} \to Q^{2}_{1})$.
 The explicit form of the quantities A is presented in {\bf appendix
C}.

\noindent The contribution of the two-photon exchange diagrams (see
 figure 2(4,4$'$; 5,5$'$)) is given by
\begin{equation}
\begin{array}{ll}
\displaystyle
 \sigma_{{2\gamma}}=&
\displaystyle
\frac{\alpha}{\pi}
\biggr(\delta_{2\gamma}^{l,u} \sigma_{o}^{l,u}
+\delta_{2\gamma}^{e,u} \sigma_{o}^{e,u}
+(\delta_{2\gamma}^{i1,u} 
+\delta_{2\gamma}^{i2,u})\sigma_{o}^{i,u}\biggl)+
   \\[0.5cm]
&\displaystyle
+\sum_{B} \sum_{T}P_{B}P_{T}\frac{\alpha}{\pi}
\biggr(\delta_{2\gamma BT }^{l,p} \sigma_{o BT}^{l,p}
+\delta_{2\gamma BT}^{e,p} \sigma_{o BT}^{e,p}+
   \\[0.5cm]
&\displaystyle
+(\delta_{2\gamma BT}^{i1,p} 
+\delta_{2\gamma BT}^{i2,p})\sigma_{o BT}^{i,p}\biggl).
\end{array}
\label{two}
\end{equation}
%where $h$$=$$l$, $e$, $i1$, $i2$.
The explicit formulae for the corrections $\delta_{2\gamma (BT)}^{h,u(p)}$
are given in the {\bf appendix D}.

\subsection{R-contribution}
In order to extract the infrared divergence, which appears when we integrate
over the real photon phase space, into a separate term we make,
 according to \cite{BS}, an identity transformation\footnote{For details about
the method used for cancellation of infrared divergence see \cite{BS}}

\begin{equation}
\sigma_{R}=\sigma_{R}-\sigma_{IR}+\sigma_{IR}=\sigma^{F}_{R}+\sigma_{IR}
\label{srf}
\end{equation}
where $\sigma_{IR}$ is infrared divergent and $\sigma^{F}_{R}$
(see {\bf appendix E}) is finite when $k \to 0$.
%After spliting the integration region over $v$ by infinitezimal
%parameter $\bar v$ we separate the infrared divergent CS
%($\sigma_{IR}$) into two parts: one infrared divergent
%(when $0 \le \bar v \le v$) and one free of divergences
%(where $v \le \bar v \le v_{max}$)
\begin{equation}
   \sigma_{IR}=
\frac{\alpha}{\pi}\delta_{soft}\sum_{k}\sigma_{o}^{k}
   + \sigma^{H}
\end{equation}
%here $k$$=${\it l, e, i}.
For the ``soft'' part we have
\begin{equation}
\delta_{soft}=\frac{1}{\pi}
\int\limits_{0}^{\bar
v}dv\int\frac{d^{n-1}k}{(2\pi\mu)^{n-1}k_{0}}F_{IR}\delta
((\Lambda-k)^{2}-m^{2})
\end{equation}
where $\mu$ is an arbitrary parameter of mass dimension, $n$ is the dimension of space, 
$\Lambda$$=$$p_{1}+k_{1}-k_{2}$ and 
$$
F_{IR}=
-\frac{m^{2}}{z_{1}^{2}}-\frac{m^{2}}{z_{2}^{2}}
+\frac{Q^{2}_{m}}{z_{1}z_{2}}
+\frac{X}{uz_{2}}+\frac{X}{uz_{1}}
-\frac{S}{vz_{2}}-\frac{S}{vz_{1}}
-\frac{m^{2}}{u^{2}}-\frac{m^{2}}{v^{2}}
+\frac{Q^{2}_{m}}{uv}.
$$
Here\; $z_{2}=2k_{2}.k$,\;$u=2p_{1}.k$\;\; 
and \;$X$$=$$S-Q^{2}-v$=$S(1-y)-v$.

\noindent The ``hard'' part is presented as
\begin{equation}
 \sigma^{H}=
\frac{\alpha}{\pi}\bar \delta\sum_{k}\sigma_{o}^{k}+
    \sigma^{H}_{1}.
\end{equation}
In this expression the first term is infrared divergent and the second
one is free of\; 
infrared divergence and is given 
explicitly by
\begin{equation}
\begin{array}{ll}
\displaystyle
 \sigma^{H}_{1}=&
\displaystyle
 \frac{4\pi\alpha^{2}S}{\lambda_{s}}
\frac{\alpha}{\pi}\int\limits_{0}^{v_{max}}dv\biggr[\biggr(
\frac{-X_{o}}{2Q^{4}}+\frac{m^{2}}{Q_{1}^{4}}
+\frac{m^{2}}{Q^{2}Q^{2}_{1}}\biggl)J(Q^{2},v)
   \\[0.5cm]
&\displaystyle
+\biggr(\frac{U_{1}}{Q^{4}}+\frac{U_{2}}{Q_{1}^{4}}
+\frac{U_{3}}{Q^{2}Q^{2}_{1}}\biggl)\Delta J(Q^{2},v)+
\sum_{B} \sum_{T}P_{B}P_{T}T_{BT}\biggl].
\end{array}
\end{equation}
Here
\begin{eqnarray}
\hspace*{3cm}
\fl T_{ll}=
-\biggr(\frac{S}{2Q^{4}}+\frac{1}{Q^{2}}
\biggl)J(Q^{2},v)
+\biggr(\frac{P_{1ll}}{Q^{4}}+\frac{P_{2ll}}{Q^{4}_{1}}+
\frac{P_{3ll}}{Q^{2}Q^{2}_{1}}\biggl)\Delta J(Q^{2},v)
\nonumber \\\displaystyle
\hspace*{3cm}
\fl  T_{tt}=
\biggr(\frac{P_{1tt}}{Q^{4}}+\frac{P_{2tt}}{Q^{4}_{1}}+
\frac{P_{3tt}}{Q^{2}Q^{2}_{1}}\biggl)\Delta J(Q^{2},v)
\nonumber \\\displaystyle
\hspace*{3cm}
\fl
T_{\bot \bot}=\biggr(\frac{2Q^{2}_{1}-v}{2Q^{2}Q^{2}_{1}}\biggl)J(Q^{2},v)+
\biggr(\frac{P_{1 \bot \bot}}{Q^{4}}+\frac{P_{2 \bot \bot}}{Q^{4}_{1}}+
\frac{P_{3 \bot \bot}}{Q^{2}Q^{2}_{1}}\biggl)\Delta J(Q^{2},v)
\nonumber \\\displaystyle
\hspace*{3cm}
\fl  T_{lt}=-\biggr(\frac{{\cal K}}{2mQ^{2}Q^{2}_{1}}\biggl)J(Q^{2},v)+
\biggr(\frac{P_{1tt}}{Q^{4}}+\frac{P_{2tt}}{Q^{4}_{1}}+
\frac{P_{3tt}}{Q^{2}Q^{2}_{1}}\biggl)\Delta J(Q^{2},v)
\nonumber \\\displaystyle
\hspace*{3cm}
\fl  T_{tl}=T_{lt} \qquad\;\;
  T_{l \bot}=T_{ \bot l}=
  T_{t \bot}=T_{ \bot t}=0
\nonumber
\end{eqnarray}
$$
\hspace*{-1cm}
\begin{array}{ll}
\displaystyle
J(Q^{2},v)=&
\displaystyle
2(Q_{m}^{2}L_{m}-1)+X(L_{X}+L_{AX})-
\displaystyle
   \\[0.5cm]
&\displaystyle
-S(L_{S}+L_{AS})+2(Q^{2}_{m}L_{Y}-1)+\frac{v}{\tau}
\end{array}
$$
\hspace*{3cm}
$\tau$=$m^{2}+v$
$$
\hspace*{-3cm}
\Delta J(Q^{2},v)=\frac{1}{v}\biggr[J(Q^{2},v)-J(Q^{2},0)\biggl].
$$
 The explicit expressions for $L_{X}$,\; $L_{AX}$,\; $L_{AS}$\;and\; 
$L_{Y}$\; may be found  in \cite{BS} (formulae (32), (33), (34) and (38) 
respectively).

At the last we present a scheme for approximate
consideration of the ``multi-soft-photon'' emission. It is achieved by
means of the so-called exponentiation procedure. 
The correction due to the consideration of  ``soft-photons'' can be
given by
\begin{equation}
\delta^{l}_{inf}=2\biggr[\biggr(\ln{\frac{Q^{2}}{m^{2}}}-1\biggl)\ln\frac{v_{max}^{2}}{SX_{o}}
+\ln{\frac{X_{o}}{m^{2}}}\ln\frac{v_{max}^{2}}{SQ^{2}}
-\ln{\frac{S}{m^{2}}}\ln\frac{v_{max}^{2}}{X_{o}Q^{2}}
\biggl]
\end{equation}
\begin{equation}
\delta^{l}_{inf}=\delta^{e}_{inf}=\delta^{i}_{inf}.
\end{equation}
(See for more details
\cite{MoT}, \cite{SLAC}-\cite{Yen}).
 This correction modifies the expression of cross section
 (\ref {csm}) in the next form:
\begin{equation}
\begin{array}{ll}
\displaystyle
\sigma+\sum_{k}\sigma^{k}_{o}\to&
\displaystyle
  \exp\biggr(\frac{\alpha}{\pi}\sum_{k}\delta^{k}_{inf}\biggl)\biggr[\sum_{k}
 \biggr(1
+ \frac{\alpha}{\pi}\sum_{f} \delta_{f}^{k}
-\frac{\alpha}{\pi}\delta^{k}_{inf}\biggl)\sigma_{o}^{k}+
\displaystyle
   \\[0.5cm]
&\displaystyle
+\sum_{k}\biggr(\sigma_{R}^{k,F}
+\sigma_{1}^{k,H}\biggl)+
\sum_{h}\biggr(\sigma_{2\gamma}^{h}+
\sigma_{amm}^{h}\biggl)
\biggl].
\end{array}
\end{equation}
In this formula  the Born
contribution has been included in $\sigma$ (\ref {csm}). The sums over $k$, $h$ and $f$ have  the 
same sense as in (\ref {csm}) and (\ref {fact}).

\section{Conclusions}
The QED lowest-order RC to the two 
identical polarized fermion scattering 
 have been
calculated
using a covariant approach. The contributions of
``hard'' and ``soft'' photons as well as two-photon exchange
have been considered exactly.     
 The ultrarelativistic approximation is used to consider the 
``multi-soft-photon'' contribution. Obtained results are free of set-up dependent
parameters (e.g. $\Delta E$).   
The analytical results have been obtained using the system of analytical calculations
 REDUCE \cite{Red}  
and are in
agreement with cross-symmetry requirements. 

 We use
polarized M{\o}ller (electron-electron) scattering at fixed target experiment to
illustrate (figure 3) how the formulae,
given above, work. Numerical calculation was perfomed using the FORTRAN
code
M{\O}LLERAD without any experimental cuts and restrictions.   
\begin{figure}[h]
\vspace{2.0cm}
\hspace{-1.0cm}
\unitlength 1mm
\begin{tabular}{cc}
\begin{picture}(40,40)
\put(10,-2){
\epsfxsize=7.0cm
\epsfysize=7.0cm
\epsfbox{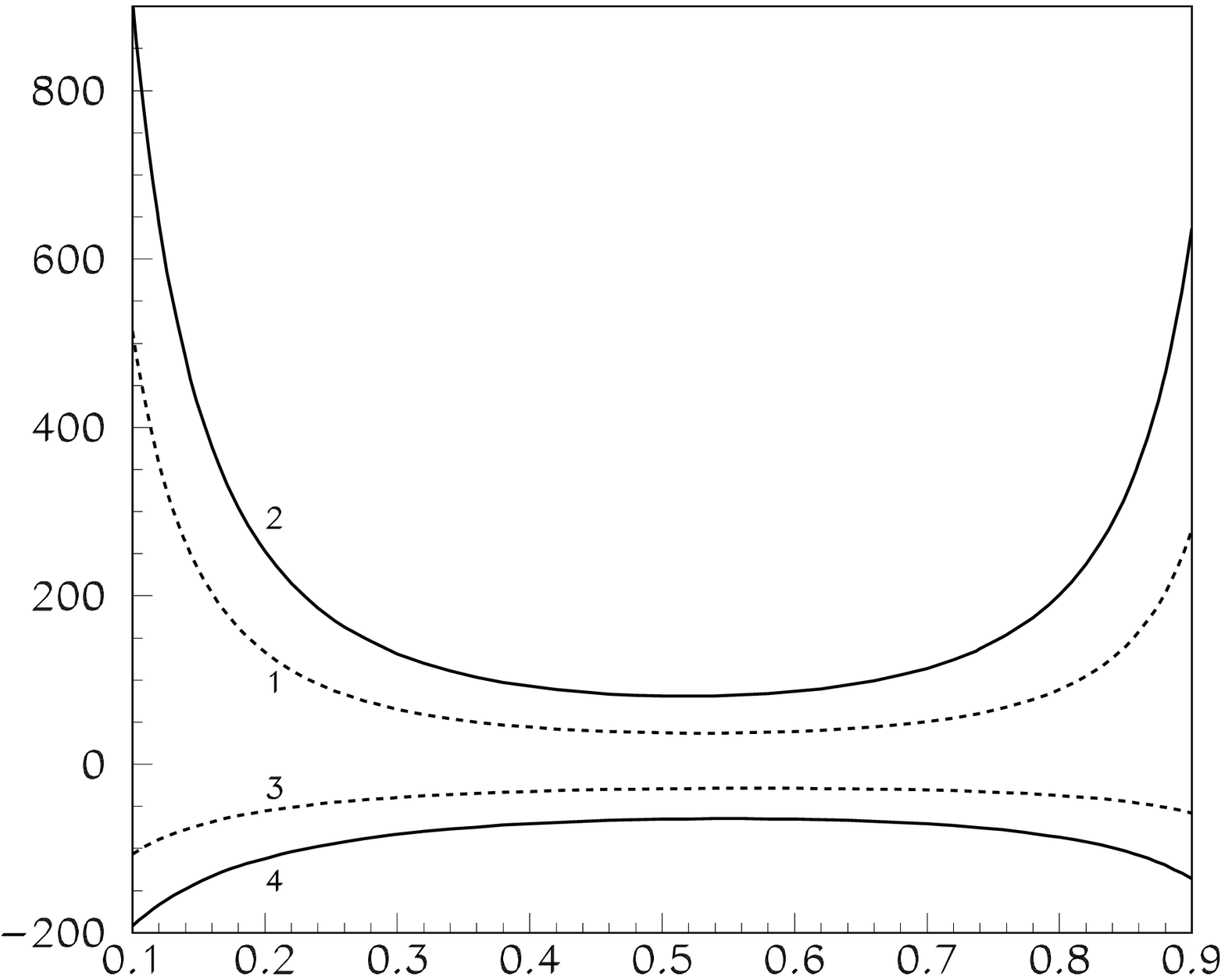}
}
\put(46.0,1.0){\makebox(0,0){\small(a)}}
\put(72.0,2.0){\makebox(0,0){\small y}}
\put(27.0,60.4){\makebox(0,0){\small
$\sigma^{p(u)}$;$\sigma^{p(u)}_{o},nb$}}
\end{picture}
\vspace{2.0cm}
\hspace{3.5cm}
\unitlength 1mm
\begin{picture}(40,40)
\put(10,-2){
\epsfxsize=7.0cm
\epsfysize=7.0cm
\epsfbox{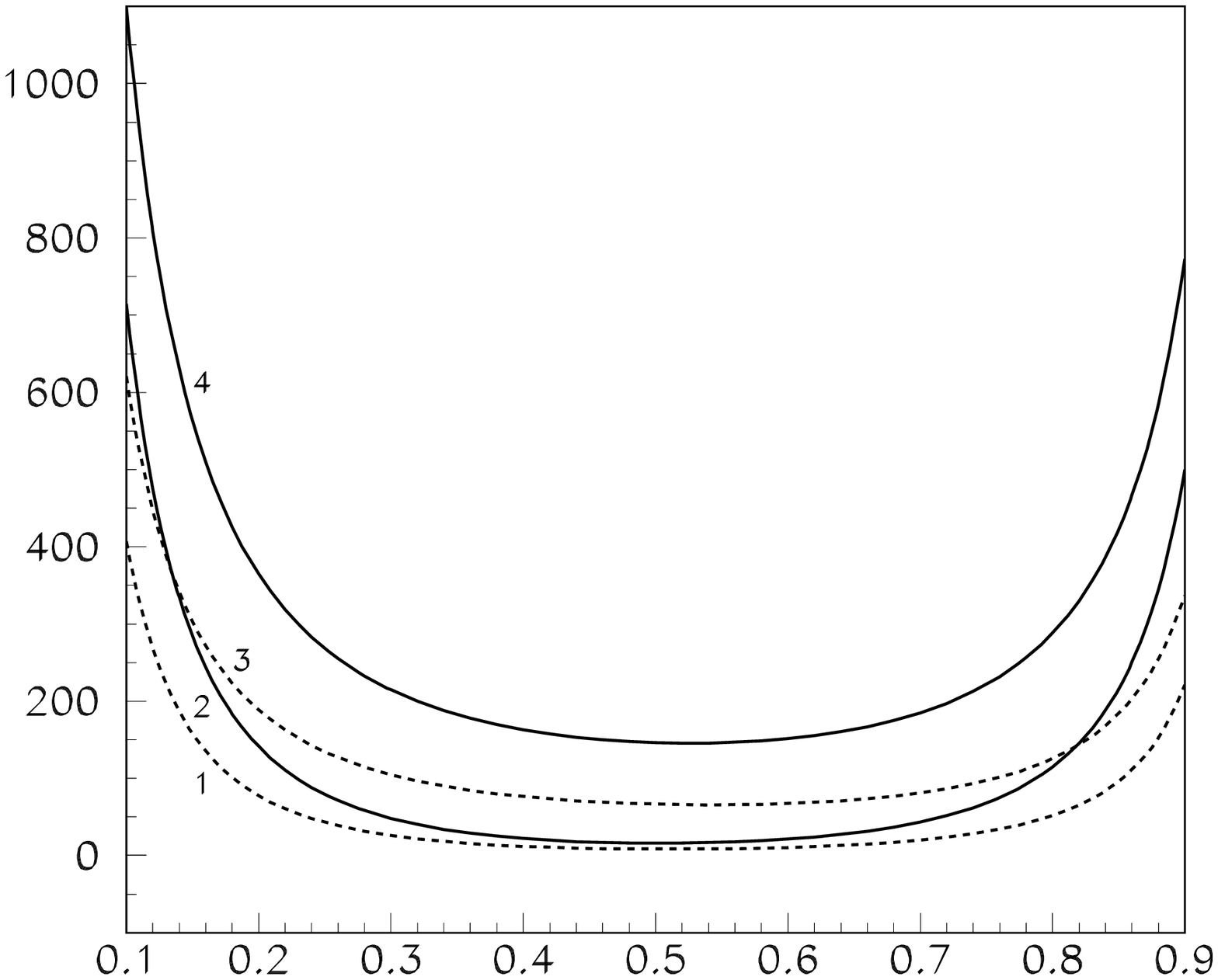}
}
\put(46.0,1.0){\makebox(0,0){\small(b)}}
\put(72.0,2.0){\makebox(0,0){\small y}}
\put(27.0,60.4){\makebox(0,0){\small $\sigma^{\uparrow
\uparrow, \uparrow \downarrow}$;$\sigma^{\uparrow \uparrow, \uparrow
\downarrow}_{o},nb$}}
\end{picture}
\end{tabular}
\end{figure}
%----CS-------------------------------
\begin{figure}[h]
\vspace{-0.50cm}
\hspace{-1.0cm}
\unitlength 1mm
\begin{tabular}{cc}
\begin{picture}(40,40)
\put(10,-2){
\epsfxsize=7.0cm
\epsfysize=7.0cm
\epsfbox{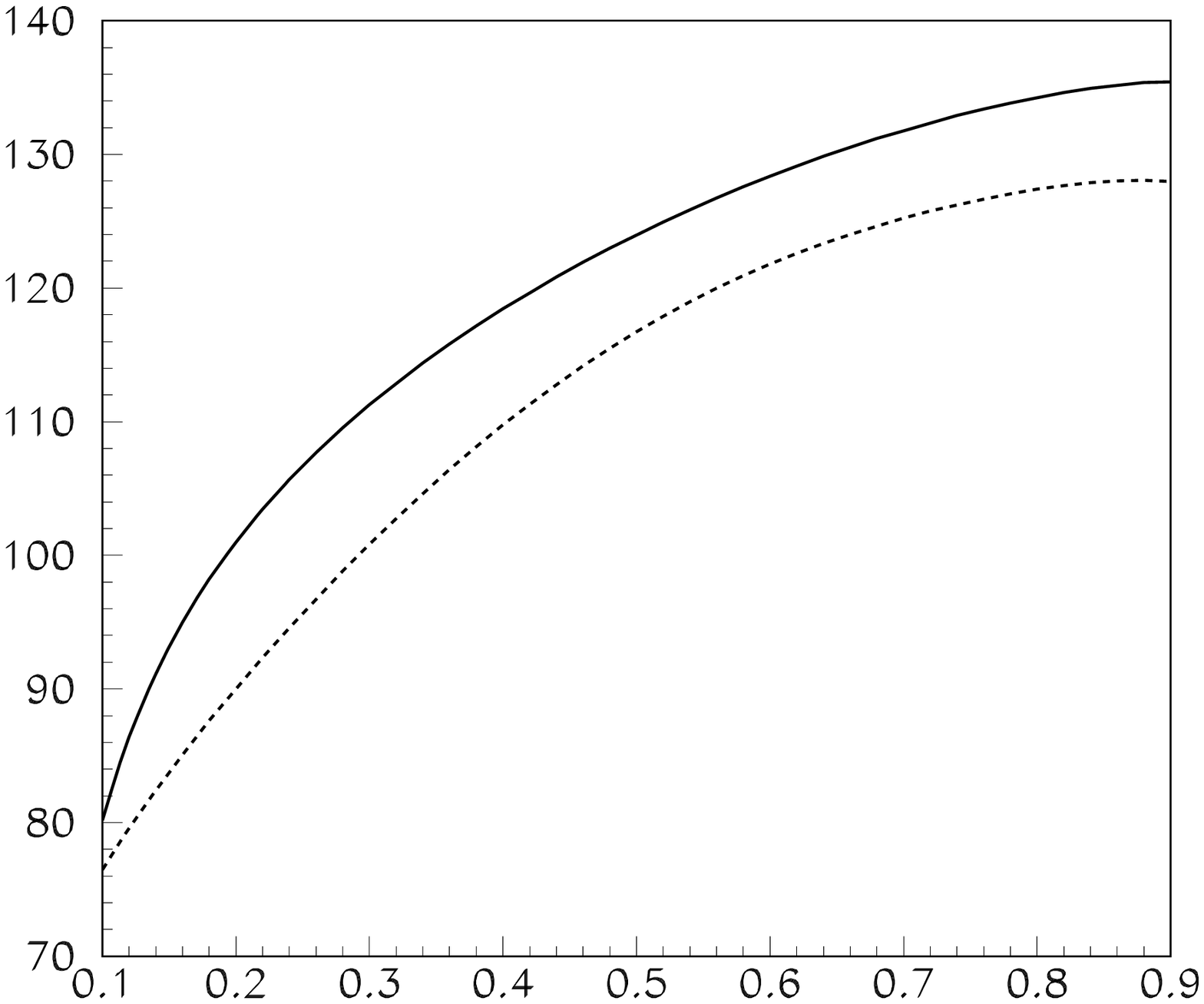}
}
\put(46.0,1.0){\makebox(0,0){\small(c)}}
\put(72.0,2.0){\makebox(0,0){\small y}}
\put(21.0,60.4){\makebox(0,0){\small $\delta^{p(u)},\%$}}
\end{picture}
\vspace{2.0cm}
\hspace{3.5cm}
\unitlength 1mm
\begin{picture}(40,40)
\put(10,-2){
\epsfxsize=7.0cm
\epsfysize=7.0cm
\epsfbox{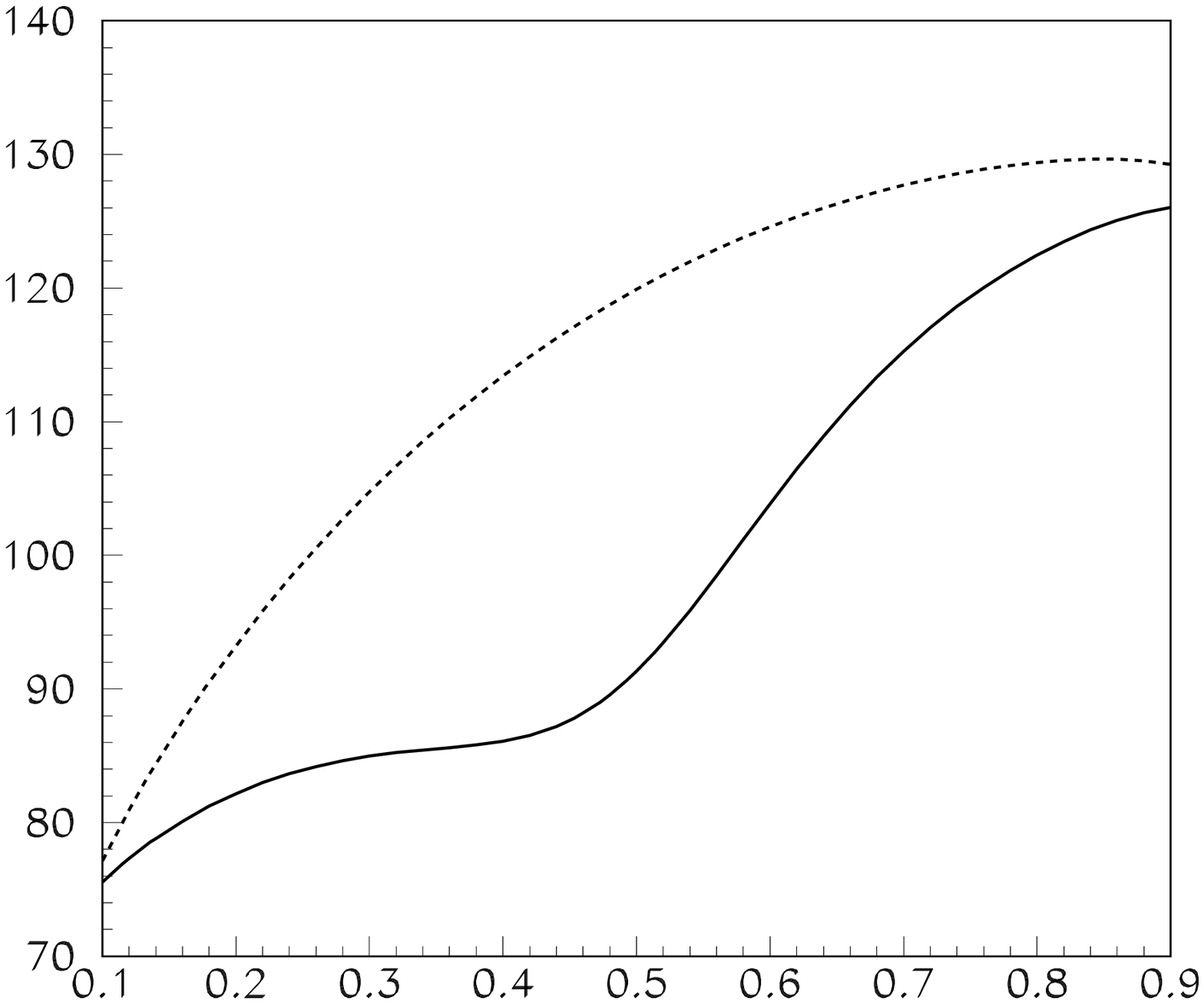}
}
\put(46.0,1.0){\makebox(0,0){\small(d)}}
\put(72.0,2.0){\makebox(0,0){\small y}}
\put(21.0,60.4){\makebox(0,0){\small $\delta^{\pm},\%$}}
\end{picture}
\end{tabular}
\label {fig}
\vspace{-2.0cm}
\caption{\protect \it y-dependence of the  
M{\o}ller scattering cross section and RC to it. (a): $\sigma_{o}^{u}$(1),
$\sigma^{u}$(2),
 $\sigma_{o}^{p}$(3), $\sigma^{p}$(4); (b): 
 $\sigma_{o}^{\uparrow \uparrow}$(1),
 $\sigma^{\uparrow \uparrow}$(2),
 $\sigma_{o}^{\uparrow \downarrow}$(3),
 $\sigma^{\uparrow \downarrow}$(4); (c):
$\delta^{p(u)}$
solid(dotted) line; (d) $\delta^{\pm}$ solid(dotted)
line.
 $E$=100 Gev. Fixed target experiment. Longitudinal
polarized beam
and target.}
\end{figure}

\noindent The quantities shown in figure 3 are defined as
follows 
$\sigma^{\uparrow \uparrow, \uparrow \downarrow}$=$\sigma_{o}\pm \sigma$
 ($P_{B}$=$P_{T}$=1, $P_{B}$=$-P_{T}$=1), and 
\begin{equation}
\sigma^{\uparrow \uparrow, \uparrow \downarrow}=\sigma^{u}\pm \sigma^{p}=
\sigma^{u}_{o}(1+\delta^{u})\pm\sigma^{p}_{o}(1+\delta^{p})=
\sigma^{\uparrow \uparrow, \uparrow \downarrow}_{o}(1+\delta^{\pm}).
\label {sigpm}
\end{equation}
\noindent Where $\sigma^{\uparrow
\uparrow,
\uparrow \downarrow}_{o}=\sigma_{o}^{u} \pm \sigma^{p}_{o}$.
 The values $\delta^{p(u)}$, 
$\delta^{\pm}$ which obey the relations       
\begin{equation}
\delta^{p(u)}=\frac{\sigma^{p(u)}}{\sigma^{p(u)}_{o}}
\end{equation}
\noindent and   
\begin{equation}
\delta^{\pm}=\frac{\sigma^{u}_{o}\delta^{u}\pm
\sigma^{p}_{o}\delta^{p}}{\sigma^{u}_{o}\pm
\sigma^{u,p}_{o}}
\end{equation}   
\noindent are the lowest-order RC to $\sigma_{o}^{p(u)}$ and
$\sigma_{o}^{\uparrow \uparrow, \uparrow \downarrow}$ respectively. 
    
 From figure 3 follows that RC 
to polarized(unpolarized) cross section $\delta^{p(u)}$ and
 ``observed'' cross 
section $\delta^{\pm}$  
are about $100\%$ \footnote{Since the RC of the lowest-order to Born cross
section
are about
100$\%$, for data analysis in a concrete set-up, either RC of the
$\alpha^{2}$-order should be investigated or 
experimental conditions are required where radiative effects could be
reduced due to experimental cuts.}. 
 We can see also, that $\delta^{p}$$\simeq$ $\delta^{u}$$\simeq$
$\delta^{\pm}$; the corrections $\delta^{u(p)}$,
$\delta^{-}$ have similar behavior and the difference, observed between
the behaviors of $\delta^{+,-}$ points out a non-vanishing(vanishing)
contribution of       
RC to the polarization asymmetry in the region where
$\delta^{-}$$>$ $\delta^{+}$($\delta^{-}$$\simeq$ $\delta^{+}$). A
detailed numerical analysis of RC to
cross section and polarization asymmetry for polarized M{\o}ller
scattering, including experimental conditions, will be a subject of separate
investigation.

The formulae proposed in this paper have been tested in a wide
kinematical range
of beam energy ($0.1$-$4000$ Gev.) at fixed target experiment. 
The obtained results may be used for numerical analysis of observables
at fixed target set-up (e.g. for M{\o}ller polarimeter) and 
colliders 
(e.g. $e^{-}e^{-}$, $\mu^{-}\mu^{-}$). However, for collideres,
proposed results can be used only at low energies, when $\sqrt{s} \ll
M_{Z_{o}}$. For 
a more accurate calculation at colliders, electroweak corrections have to be
considered.

\section*{Acknowledgements}
The authors are grateful to I. Akushevich, V. Mossolov, P. Kuzhir and A.
Tolkachev for fruitful discussions and comments. 
\section*{Appendix A}
In this appendix the quantities (see (\ref {born})), which are necessary to
calculate 
the Born cross section are shown.          
\begin{eqnarray}
\label{csec}
\fl U_{1}=\frac{1}{2}\biggr(S^{2}+Q_{1}^{4}\biggl)+2m^{2}(S-2Q^{2}-m^{2})
\nonumber
 \\
\fl U_{2}=\frac{1}{2}\biggr(S^{2}+Q^{4}\biggl)+2m^{2}(S-2Q_{1}^{2}-m^{2})
\nonumber
\\
\fl U_{3}=S(S-4m^{2})
\nonumber
 \\
\fl P_{1ll}=Q^{2}\biggr[\frac{Q^{2}d_{s}^{2}}{2\lambda_{s}}-S\biggl]
\nonumber
\\
\fl
P_{2ll}=Q^{2}\biggr[\frac{Q^{2}d_{s}^{2}}{2\lambda_{s}}-2m^{2}\biggl]+\frac{1}{2}\lambda_{s}-
Sd_{s}
\nonumber
\\
\fl P_{3ll}=-\frac{1}{\lambda_{s}}\biggr[S^{4}+4m^{2}(-S^{2}Q^{2}+m^{2}(
-S^{2}+2Q^{4}+4m^{2}Q^{2}))\biggl]
\nonumber
\end{eqnarray}
\begin{eqnarray}
\label{bornpa}
\fl P_{1tt}=P_{2tt}=\frac{2m^{2}Q^{4}}{ {\cal K}^{2} \lambda_{s}}
\biggr(4m^{2}(4m^{4}Q^{2}+m^{2}(Q^{4}-2SX_{o})-SQ^{2}X_{o})+S^{2}X_{o}^{2}\biggl)
%\nonumber
\\
\fl P_{3tt}=-\frac{SQ^{4}}{{\cal K}^{2} \lambda_{s}}
\biggr(4m^{2}(4m^{4}Q^{2}+m^{2}(Q^{4}-2SX_{o})-SQ^{2}X_{o})+S^{2}X_{o}^{2}\biggl)
\nonumber
\end{eqnarray}
\begin{eqnarray}
\fl P_{1 \bot \bot}=-2m^{2}Q^{2}&\qquad
P_{2 \bot \bot}=2m^{2}(2m^{2}-X_{o})&\qquad
P_{3 \bot \bot}=2m^{2}(-2S+X_{o})+X_{o}Q^{2}
\nonumber
\end{eqnarray}
\begin{eqnarray}
\fl P_{1lt}=P_{1tl}=\frac{mQ^{4}}{ {\cal K} \lambda_{s}}
\biggr(2m^{2}(2m^{2}(2m^{2}-2S-X_{o})+S(S-2X_{o})-S^{2}X_{o}\biggl)
\nonumber
\\
\fl P_{2lt}=P_{2tl}=\frac{mQ^{2}}{ {\cal K} \lambda_{s}}
\biggr(2m^{2}Q^{2}(2m^{2}(2m^{2}-2S-X_{o})+S(S-2X_{o})+
\nonumber \\
\hspace*{-1cm}
+\lambda_{s}(2m^{2}(Q^{2}-2m^{2})+SX_{o})-S^{2}X_{o}Q^{2})\biggl)
\nonumber
\\
\fl P_{3lt}=P_{3tl}=-\frac{SQ^{4}}{{\cal K} \lambda_{s}}
\biggr(4m^{2}(4m^{4}Q^{2}+m^{2}(Q^{4}-2SX_{o})-SQ^{2}X_{o})+S^{2}X_{o}^{2}\biggl)
\nonumber
\end{eqnarray}
\begin{eqnarray}
\fl P_{1l \bot}=P_{1 \bot l}=P_{1t \bot}=P_{1 \bot t}=0\qquad\qquad
P_{2l \bot}=P_{2 \bot l}=P_{2t \bot}=P_{2 \bot t}=0\;\;\;
\nonumber
\\
\fl P_{3l \bot}=P_{3 \bot l}=P_{3t \bot}=P_{3 \bot t}=0
\nonumber
\end{eqnarray}
\noindent In (\ref {bornpa}) $Q^{2}_{1}$=$X_{o}-2m^{2}$,\;$d_{s}$=$S+2m^{2}$. 

\section*{Appendix B}

Here we present a set of integrals, which are necessary for calculations of
the
contribution
$\sigma_{R}^{F}$ (see (\ref {srf})). This set is complementary to the
one that was given in \cite{BS, Prep2}.

The integrals have the form
\begin{equation}
\biggr[{\huge A}\biggl]=\int
\limits_{t_{min}}^{t_{max}}dt\int
\limits_{z_{2}^{min}}^{z_{2}^{max}}\frac{dz_{2}}{\sqrt{R_{z_{2}}}}{\huge
A}.
\end{equation}
\noindent The limits of kinematical variables $t$ and $z_{2}$ are
$t_{max/min}=\displaystyle\frac{vS_{x}+2m^{2}Q^{2}\pm
v\sqrt{\lambda_{Y}}}{2\tau}$,\;\;
 $z^{max(min)}_{2}$ were given first in \cite{Prep3} (See also
\cite{Aku}). $R_{z_{2}}$ is the Gram determinant.  
\begin{eqnarray}
\fl \biggr[\frac{t^{2}}{z_{2}}\biggl]=\frac{1}{2A}\biggr[\frac{v}{\tau^{2}}
(ER_{ex}+F\tau)-\frac{3BEv}{A\tau}+\biggr(\frac{3B^{2}}{A}-C\biggl)L_{A}
\biggl] 
\\
\fl \biggr[\frac{t^{2}}{z_{1}}\biggl]=-\biggr[\frac{t^{2}}{z_{2}}\biggl]
(S \to-X)
%\nonumber
 \\
\fl \biggr[\frac{t}{z_{2}^2}\biggl]=\frac{1}{A}\biggr[\frac{1}{Mv}
\biggr(\frac{Q^{2}SW(Q^{2}+X)}{m^{2}}-EF\biggl)+EL_{A}
\biggl]
 \\
\fl \biggr[\frac{t}{z_{1}^{2}}\biggl]=\biggr[\frac{t}{z_{2}^{2}}\biggl]
(S \to-X)
%\nonumber
\\
\fl \biggr[\frac{t^{2}}{z_{2}^2}\biggl]=
\frac{1}{A^{2}}\biggr[\frac{1}{Mv}
\biggr(E(AH+2BF)- 2WEQ^{2}Sv-
\nonumber \\
\lo -W(X+Q^{2})\biggr(Fv+
\frac{BQ^{2}S}{m^{2}}\biggl)\biggl)+(AF-3BE)L_{A}
\biggl]
 \\
\fl \biggr[\frac{t^{2}}{z_{1}^{2}}\biggl]=\biggr[\frac{t^{2}}{z_{2}^{2}}\biggl]
(S \to-X)
%\nonumber
 \\
\fl \biggr[\frac{1}{z_{2}(t-R)}\biggl]=L_{R}=
\frac{1}{SX}\ln
\frac{(t_{2}-R)[SX+(X+Q^{2})(t_{1}-R)+ \sqrt{-C(t_{1})}]} 
{(t_{1}-R)[SX+(X+Q^{2})(t_{2}-R)+\sqrt{-C(t_{2})}]}
\\
\fl t_{2}=t_{max},t_{1}=t_{min} 
\nonumber
\\
\fl \biggr[\frac{1}{z_{2}^2(t-R)}\biggl]=\frac{1}{S^{2}X^{2}}\biggr[
 F_{R}L_{R}-\frac{B_{R}}{m^{2}v}\biggl]
\\
\fl \biggr[\frac{1}{z_{2}(t-R)^2}\biggl]=\frac{1}{S^{2}X^{2}}\biggr[
\frac{vF_{R}}{X_{12}}-B_{R}L_{R}\biggl]
\\
\fl \biggr[\frac{1}{z_{2}^{2}(t-R)^2}\biggl]=\frac{1}{S^{2}X^{2}}\biggr[
L_{R}(E-\frac{3F_{R}B_{R}}{S^{2}X^{2}})+\frac{A}{m^{2}v}
+\frac{v}{X_{12}}\biggr(\frac{3F_{R}^{2}}{S^{2}X^{2}}-2\lambda_{Y}
\biggl)\biggl]
\\
\fl \biggr[\frac{1}{z_{2}(t-R)^3}\biggl]=\frac{1}{2S^{2}X^{2}}\biggr[
\frac{v}{X_{12}}\biggr(\frac{\Theta}{X_{12}}-\frac{3F_{R}B_{R}}{S^{2}X^{2}})
+L_{R}(\frac{3B_{R}^{2}}{S^{2}X^{2}}-A)
\biggl]
\\
\fl \biggr[\frac{1}{(z_{1}-P)}\biggl]=L_{P}=
\frac{1}{\hat A}\ln
\frac{X_{o}t_{2}-Sv+\sqrt{-\hat N(t_{2})}} 
{X_{o}t_{1}-Sv+\sqrt{-\hat N(t_{1})}} 
\\
\fl \biggr[\frac{t}{(z_{1}-P)}\biggl]=\frac{1}{\hat A}
\biggr[\frac{\hat Ev}{\tau}-\hat B_{P}L_{P}\biggl]
\\
\fl \biggr[\frac{1}{t(z_{1}-P)}\biggl]=L_{N}=
\frac{1}{Sv}\ln
\frac{t_{2}[Sv-X_{o}t_{1}+ \sqrt{-\hat N(t_{1})}]} 
{t_{1}[Sv-X_{o}t_{2}+ \sqrt{-\hat N(t_{2})}]} 
\\
\fl \biggr[\frac{1}{u(z_{1}-P)}\biggl]=L_{\xi}=
\frac{1}{XQ^{2}}\ln
\frac{(t_{1}-S_{x})[XQ^{2}+X_{o}(t_{2}-S_{x})+ \sqrt{-\hat N(t_{2})}]} 
{(t_{2}-S_{x})[XQ^{2}+X_{o}(t_{1}-S_{x})+ \sqrt{-\hat N(t_{1})}]} 
\\
\fl \biggr[\frac{1}{u^{2}(z_{1}-P)}\biggl]=\frac{1}{X^{2}
Q^{4}}
\biggr[\frac{\hat F_{\xi}v}{\xi_{12}}-\hat B_{\xi}L_{\xi}\biggl]
\\
\fl \biggr[\frac{1}{(z_{1}-P)^{2}}\biggl]=\frac{1}{X_{o}\hat
N_{12}}
\biggr[\hat F+X_o(\lambda_{Y}+ \hat E)
\biggl]
\\
\fl \biggr[\frac{t}{(z_{1}-P)^{2}}\biggl]=\frac{1}{\hat
A}
\biggr[\frac{1}{\hat
N_{12}}\biggr((X+v)Wv-X_{o}(\hat E^{2}+\lambda_{Y}(\hat 
E-\hat A))+
\nonumber \\
\hspace*{.5cm} +2\lambda_{Y}\hat B_{P} 
-\hat E\hat F\biggl)+ \hat E
L_{P}\biggl]
\\
\fl \biggr[\frac{1}{u(z_{1}-P)^{2}}\biggl]=\frac{1}{X^{2}
Q^{4}}
\biggr[\frac{1}{\hat N_{12}}\biggr(
(X-Q^{2})(\lambda_{Y}\hat A+\hat E^{2}-Wv)
+\hat E\hat F_{\xi}\biggl)+ \hat F_{\xi}
L_{\xi}\biggl]
\\
\fl \biggr[\frac{1}{u^{2}(z_{1}-P)^{2}}\biggl]=\frac{1}{X^{2}
Q^{4}}
\biggr[\frac{v}{\xi_{12}}
\biggr(\frac{3\hat F_{PR}^{2}}{S^{2}X^{2}}
-2 \lambda_{Y}\biggl)+ \frac{1}{\hat N_{12}}\biggr(Wv
- \hat A\lambda_{Y}\biggl)+
L_{\xi} 
(\hat E-\frac{3\hat B_{\xi}\hat F_{\xi}}{X^{2}Q^{4}})\biggl]
\\
\fl \biggr[\frac{1}{(z_{1}-P)(t-R)}\biggl]=L_{PR}=
\frac{1}{SX}\ln
\frac{(t_{2}-R)[SX+X_{o}(t_{1}-R)+ \sqrt{-\hat N(t_{1})}]} 
{(t_{1}-R)[SX+X_{o}(t_{2}-R)+\sqrt{-\hat N(t_{2})}]} 
\\
\fl \biggr[\frac{1}{(z_{1}-P)(t-R)^{2}}\biggl]=\frac{1}{S^{2}
X^{2}}\biggr[
\frac{v\hat F_{PR}}{X_{12}}-\hat B_{PR}L_{PR}\biggl]
\\
\fl \biggr[\frac{1}{(z_{1}-P)(t-R)^{3}}\biggl]=\frac{1}{2S^{2}
X^{2}}\biggr[
\frac{v}{X_{12}}\biggr(\frac{\Sigma}{X_{12}}-\frac{3\hat
F_{PR} \hat B_{PR}}{S^{2}X^{2}})
+L_{PR}(\frac{3B_{PR}^{2}}{S^{2}X^{2}}-\hat A)
\biggl]
\\
\fl \biggr[\frac{1}{(z_{1}-P)^{2}(t-R)}\biggl]=\frac{1}{S^{2}X^{2}}
\biggr[\frac{1}{\hat
N_{12}}\biggr(Q^{2}(\lambda_{Y}\hat A+\hat E^{2}-Wv)
-\hat E\hat F_{PR}\biggl)+ \hat F_{PR}
L_{PR}\biggl]
\\
\fl \biggr[\frac{1}{(z_{1}-P)^{2}(t-R)^{2}}\biggl]=\frac{1}{
S^{2}X^{2}}
\biggr[\frac{v}{X_{12}}
\biggr(\frac{3\hat F_{PR}^{2}}{S^{2}X^{2}}
-2 \lambda_{Y}\biggl)+
 \frac{1}{\hat N_{12}}\biggr(Wv
-\hat A\lambda_{Y}
\biggl)+
\nonumber \\
\hspace*{2.5cm} 
+L_{PR}
(\hat E-\frac{3\hat B_{PR}\hat F_{PR}}{S^{2}X^{2}})\biggl].
\end{eqnarray}

\begin{eqnarray}
\fl A=(X+Q^{2})^{2}-4m^{2}\tau\;\;\;\;&
\hat A=(S-Q^{2})^{2}-4m^{2}\tau\;\;\;\;
\nonumber \\\displaystyle
\fl  B=vB_{o}-Q^{2}\lambda_{s}&
\hat  B=v\hat B_{o}-Q^{2}\lambda_{x}&
\nonumber \\\displaystyle
\fl B_{o}=SQ^{2}+2m^{2}S_{x}&
\hat B_{o}=-XQ^{2}+2m^{2}S_{x}&
\nonumber \\\displaystyle
\fl C =Q^{2}\lambda_{s}&
\hat C =Q^{2}\lambda_{x}&
  \\\displaystyle
\fl  E=vX-Q^{2}d_{s}&
\hat  E=vS-Q^{2}(X-2m^{2})&
\nonumber \\\displaystyle
\fl F=Q^{2}F_{o}&
\hat F=Q^{2}\hat F_{o}&
\nonumber \\\displaystyle
\fl F_{o}=SS_{x}+2m^{2}Q^{2}&
\hat F_{o}=XS_{x}-2m^{2}Q^{2}&
\nonumber
\end{eqnarray}
\begin{eqnarray}
\fl M=Q^{2}(SX-Q^{2}m^{2})\;\;\;\;&
 H=EQ^{4}m^{2}+FR_{ex}\;\;\;\;&
W=M-m^{2}\lambda_{Y}
\nonumber \\\displaystyle
\fl \hat B_{PR}=SX(X+v)&
\hat H=\hat EQ^{4}m^{2}+\hat FR_{ex}&
R_{ex}= S_{x}v+2m^{2}Q^{2}
\nonumber \\\displaystyle
\fl B_{R}=SX(X+Q^{2})&
F_{R}=SX(v-Q^{2})&
\hat B_{P}=-Sv(X+v)
 \\\displaystyle
\fl \hat F_{P}=-Sv(v+Q^{2})&
 \hat F_{PR}=F_{R}&
\hat B_{\xi}=XQ^{2}(X+v)
\nonumber \\\displaystyle
\fl P=X+v-2m^{2}\;\;\;\;&
S_{x}=S-X\;\;\;\;&
 \lambda_{Y}=(Q^{2}+v)^{2}+4m^{2}Q^{2}
\nonumber
\end{eqnarray}
\begin{eqnarray}
%\nonumber \\\displaystyle
\fl \hat F_{\xi}=-XQ^{2}(v+Q^{2})\;\;\;\;\;\;
\Sigma=SXv(SQ^{2}-Xv)\;\;\;\;
 R=S-2m^{2}
\nonumber \\\displaystyle
\fl t_{1}t_{2}=\frac{m^{2}Q^{4}}{\tau}\qquad\qquad\;\;\;
 \Theta=SXv(XQ^{2}-Sv)\;\;\;\;
 X_{12}=SXv
  \\\displaystyle
\fl \xi_{12}=m^{2}v^{2}\qquad\qquad\;\;\;\;\;
 \hat N_{12}=5SXQ^{2}v+2(X^{2}Q^{4}+S^{2}v^{2})
\nonumber
\end{eqnarray}
\begin{eqnarray}
\fl \sqrt{-C(t_{1(2)})}=\frac{Et_{1(2)}+F}{\sqrt{\lambda_{Y}}}&
\nonumber  \\\displaystyle
\fl \sqrt{-\hat C(t_{1(2)})}=\frac{\hat Et_{1(2)}+\hat F}{\sqrt{\lambda_{Y}}}&
  \\\displaystyle
\fl \sqrt{-\hat N(t_{1(2)})}=\sqrt{-\hat C(t_{1(2)})}+X_{o}\sqrt{\lambda_{Y}}.
\nonumber
\end{eqnarray}

\section*{Appendix C}
The quantities A (see (\ref{amm})) are given by
\begin{eqnarray}
\label {ammt}
\fl A_{1}=2(Q^{2}-2m^{2})
\nonumber \\
\fl A_{2}=-2m^{2}(3S+X_{o})
 \nonumber 
\end{eqnarray}
\begin{eqnarray}
\fl A_{3}=2\biggr(8m^{4}-6m^{2}X_{o}+X_{o}^{2}\biggl)
\nonumber \\
\fl A_{4}=2(2S^{2}+X_{o}^{2}-2m^{2}(3S+Q^{2}))
\nonumber \\
\fl A_{5ll}=\frac{Q^{2}}{\lambda_{s}}\biggr(4m^{4}+4m^{2}S+S^{2}\biggl)-2S-Q^{2}
\nonumber \\
\fl
A_{6ll}=m^{2}\biggr(1-\frac{SQ^{2}}{\lambda_{s}}\biggl)
+2\biggr(-\frac{2S^{2}Q^{2}}{\lambda_{s}}+3Q^{2}-2S\biggl)+
\\
\hspace*{-1.4cm}
+\frac{1}{m^{2}}\biggr(-2\lambda_{s}-\frac{S^{3}Q^{2}}{\lambda_{s}}+2S^{2}+SQ^{2}\biggl)
\nonumber
 \\
\fl
A_{7ll}=\frac{2m^{4}}{\lambda_{s}}\biggr(Q^{4}-2SX_{o}\biggl)-2m^{4}+2m^{2}\biggr(2S-Q^{2}-
\frac{SQ^{2}X_{o}}{\lambda_{s}}\biggl)+X_{o}\biggr(\frac{S^{2}X_{o}}{\lambda_{s}}-X_{o}-2S\biggl)
\nonumber \\
\fl
A_{8ll}=-\frac{4m^{4}Q^{2}}{\lambda_{s}}\biggr(2S+Q^{2}\biggl)
+2m^{2}\biggr(5Q^{2}+2S-\frac{3SQ^{4}}{\lambda_{s}}\biggl)+
\nonumber \\
\hspace*{-1cm}
+\biggr(\frac{S^{2}Q^{2}}{\lambda_{s}}(2S-3Q^{2})-2S(S+Q^{2})+5Q^{4}\biggl)+
\frac{SQ^{2}}{2m^{2}}\biggr(\frac{S^{3}}{\lambda_{s}}-\frac{S^{2}Q^{2}}{\lambda_{s}}-S+Q^{2}\biggl)
\nonumber \\
\fl A_{5tt}=\frac{2m^{2}}{{\cal
K}^{2}\lambda_{s}}\biggr(4m^{2}(4m^{6}+4m^{4}Q^{2}+m^{2}(Q^{4}-2SX_{o})
-SQ^{2}X_{o})+S^{2}X^{2}_{o}\biggl)
\nonumber \\
\fl A_{6tt}=-\frac{2Q^{2}S}{{\cal
K}^{2}\lambda_{s}}\biggr(4m^{2}(4m^{6}+4m^{4}Q^{2}+m^{2}(Q^{4}-2SX_{o})
-SQ^{2}X_{o})+S^{2}X^{2}_{o}\biggl)
\nonumber \\
\fl A_{7tt}=\frac{2m^{2}Q^{4}}{{\cal
K}^{2}\lambda_{s}}\biggr(4m^{2}(4m^{6}+4m^{4}Q^{2}+m^{2}(Q^{4}-2SX_{o})
-SQ^{2}X_{o})+S^{2}X^{2}_{o}\biggl)
\nonumber \\
\fl A_{8tt}=\frac{2m^{2}Q^{4}}{{\cal
K}^{2}\lambda_{s}}\biggr(4m^{2}(12m^{6}Q^{2}+2m^{4}(-2SX_{o}+3Q^{2})+
m^{2}Q^{2}(Q^{4}-4SX_{o}))+
\nonumber \\
+2m^{2}S(X_{o}(S-2Q^{2})(S+Q^{2}))
+Q^{2}S^{2}X^{2}_{o}\biggl).
\nonumber \\
\fl A_{5 \bot \bot}=-8mQ^{2}(m^{2}+Q^{2})
\nonumber \\
\fl A_{6 \bot
\bot}=2mX_{o}(X_{o}-8m^{2})
\nonumber \\
\fl A_{7 \bot \bot}=4m(-2m^{2}(S+2Q^{2})+S(S+Q^{2})-2Q^{4})
\nonumber \\
\fl A_{8 \bot \bot}=4mQ^{2}(3S-2Q^{2}-4m^{2}).
\nonumber
\end{eqnarray}
All details about the calculation of the quantities given above can be
found 
at htpp://www.hep.by/mollerad.htm
 
\section*{Appendix D}
In this appendix we present the explicit formulae for the two-photon
exchange contribution (see (\ref{two})). The corrections $\delta^{h,u(p)}_{2\gamma(BT)}$
have the form
\begin{equation}
\begin{array}{ll}
\displaystyle
\delta_{2\gamma}^{l,u}=&
\displaystyle
-\frac{Q^{2}}{U_{1}}\biggr[\frac{S}{2Q^{2}}\biggr(U_{1}+S^{2}\biggl)K_{s}+
\frac{X_{o}}{2Q^{2}}\biggr(U_{1}+X_{o}^{2}\biggl)K_{x_{o}}-
   \\[0.5cm]
&\displaystyle
-\frac{1}{4Q_{1}^{2}}
\biggr(Q^{2}Q_{1}^{4}+X_{o}\lambda_{x_{o}}\biggl)L_{x_{o}}
+\frac{1}{4d_{s}}
\biggr(Q^{2}d_{s}^{2}-S\lambda_{s}\biggl)L_{s}-
   \\[0.5cm]
&\displaystyle
-(S+X_{o})\biggr(4Q_{m}^{2}G_{o}^{m}+8m^{2}
g_{1}^{m}+\frac{1}{2}\ln\frac{m^{2}}{Q^{2}}\biggl)\biggl]
\label{twodel1}
%\nonumber
\end{array}
\end{equation}
$$
\hspace*{.6cm}
\Re=
\biggr(\frac{\lambda_{s}Q^{2}}{4\Delta}-S\biggl)K_{s}-
\biggr(\frac{\lambda_{x_{o}}Q^{2}}{4\Delta}+X_{o}\biggl)K_{x_{o}}-
\frac{2Q^{4}(S+X_{o})}{\Delta}G_{o}^{m}
$$
\begin{equation}
\begin{array}{ll}
\displaystyle
\delta_{2\gamma}^{i1,u}=&
\displaystyle
\Re -\frac{Q^{2}m^{2}}{U_{3}}
\biggr[S^{2}((S+Q^{2})\hat a_{1}-X_{o}a_{1})+S^{2}(\hat b_{o}+b_{o})+
4S^{2}b_{o}-
   \\[0.5cm]
&\displaystyle
-2m^{2}Q^{2}(\hat b_{1}-b_{1})+m^{2}(S+X_{o})(\hat
b_{4}+b_{4})+2m^{2}(S+Q^{2})\hat b_{4}\biggl]
\end{array}
\end{equation}

\begin{equation}
\begin{array}{ll}
\displaystyle
\delta_{2\gamma ll}^{l,p}=&
\displaystyle
\Re +\frac{2Q^{2}m^{2}}{P_{1ll}}
\biggr[Q^{2}(S+X_{o})(S\hat a_{1}+X_{o}a_{1})+
   \\[0.5cm]
&\displaystyle
+(S^{2}+X_{o}^{2})(\hat b_{o}+b_{o})+
S(X_{o}\hat b_{4}-Sb_{4})\biggl]
\end{array}
\end{equation}

\begin{equation}
\begin{array}{ll}
\displaystyle
\delta_{2\gamma tt}^{l,p}=&
\displaystyle
\Re +\frac{Q^{2}m^{2}}{P_{1tt}}
\biggr[X_{o}Q^{2}((-3S+2Q^{2})\hat a_{1}-4Sa_{1})+
   \\[0.5cm]
&\displaystyle
+4Q^{2}X_{o}(\hat b_{1}-b_{1})+
6X_{o}Q^{2}(\hat b_{4}-b_{4})\biggl]
\end{array}
\end{equation}

\begin{equation}
\begin{array}{ll}
\displaystyle
\delta_{2\gamma \bot \bot}^{l,p}=&
\displaystyle
\Re +\frac{Q^{2}m^{4}}{P_{1 \bot \bot}}
\biggr[X_{o}(S \hat a_{1}+Q^{2}a_{1})
-2S(\hat b_{o}+b_{o})+
   \\[0.5cm]
&\displaystyle
+(\frac{3}{4}Q^{2}-S)(\hat b_{1}-b_{1})-
\frac{1}{2}X_{o}(\hat b_{4}-b_{4})\biggl]
\end{array}
\end{equation}

\begin{equation}
\begin{array}{ll}
\displaystyle
\delta_{2\gamma ll}^{i1,p}=&
\displaystyle
\Re +\frac{8Q^{2}S^{2}m^{2}}{P_{3ll}}
\biggr[S\hat a_{1}+X_{o}a_{1}+\hat b_{o}+b_{o}-
b_{4}\biggl]
\end{array}
\end{equation}

\begin{equation}
\begin{array}{ll}
\displaystyle
\delta_{2\gamma tt}^{i1,p}=&
\displaystyle
\Re +\frac{Q^{2}m^{2}}{P_{3tt}}
\biggr[Q^{4}(2S\hat a_{1}+X_{o}a_{1})+SX_{o}(\hat b_{1}+b_{1})-
3SQ^{2}b_{4}\biggl]
\end{array}
\end{equation}

\begin{equation}
\begin{array}{ll}
\displaystyle
\delta_{2\gamma \bot \bot}^{i1,p}=&
\displaystyle
\Re +\frac{Q^{2}m^{4}}{P_{3 \bot \bot}}
\biggr[X_{o}^{2}\hat a_{1}+(S+Q^{2})a_{1}+
12X_{o}(\hat b_{o}+b_{o})+
   \\[0.5cm]
&\displaystyle
+S(\hat
b_{1}+b_{1})-
2Q^{2}(\hat b_{4}+b_{4})\biggl].
\end{array}
\end{equation}
The rest ones are obtained from the corrections given
above, using cross-symmetry:
\begin{eqnarray} 
\fl \delta_{2\gamma}^{e,u}=
\delta_{2\gamma}^{l,u}(\Im_{1})\qquad&
\delta_{2\gamma}^{i2,u}=
\delta_{2\gamma}^{i1,u}
(\Im_{2})
\nonumber \\\displaystyle
\fl \delta_{2\gamma ll}^{e,p}=
\delta_{2\gamma ll}^{l,p}
(P_{1ll} \to P_{2ll},\;\Im_{2})\qquad&
\delta_{2\gamma ll}^{i2,p}=
\delta_{2\gamma ll}^{i1,p}
(\Im_{2})	
\nonumber \\\displaystyle
\fl \delta_{2\gamma tt}^{e,p}=
\delta_{2\gamma tt}^{l,p}
(P_{1tt} \to P_{2tt},\;\Im_{2})\qquad&
 \delta_{2\gamma tt}^{i2,p}=
\delta_{2\gamma tt}^{i1,p}
(\Im_{2})
 \\\displaystyle
\label {twodel2}
\fl \delta_{2\gamma \bot \bot}^{e,p}=
\delta_{2\gamma \bot \bot}^{l,p}
(P_{1 \bot \bot} \to P_{2 \bot \bot},\;\Im_{2})\qquad&
\delta_{2\gamma \bot \bot}^{i2,p}=
\delta_{2\gamma \bot \bot}^{i1,p}
(\Im_{2}).
\nonumber 
\end{eqnarray}
%\hspace*{-1cm}
\noindent Here $\Im_{1}$ are the following replacements:
$$
\begin{array}{ll}
U_{1} \to U_{2},\;
Q^{2} \to Q_{1}^{2},\;
X_{o} \to Q^{2}_{m},\;
\lambda_{x_{o}} \to \lambda_{m},\;
L_{x_{o}} \to L_{m},\;
   \\[0.5cm]
%&\displaystyle
K_{s} \to \tilde K_{s},\;
 K_{x_{o}}\to \tilde K_{x_{o}},\;
  G_{o}\to\tilde G_{o},\; 
 g^{m}_{1} \to \tilde g^{m}_{1}
\end{array}
$$
and $\Im_{2}$ are the replacements: 
$$
\begin{array}{ll}
 \Delta \to \tilde \Delta,\;
Q^{2} \to Q_{1}^{2},\;
X_{o} \to Q_{m}^{2},\;
 \lambda_{x_{o}} \to \lambda_{m},\;
  G_{o}\to \tilde G_{o},\; 
   \\[0.5cm]
%&\displaystyle
 K_{s} \to \tilde K_{s},\;
  K_{x_{o}}\to \tilde K_{x_{o}},\;
 \hat a_{1} \to \hat  a_{11},\;
 a_{1} \to  a_{11},\;
 \hat b_{o} \to \hat b_{o1},\;
   \\[0.5cm]
%&\displaystyle
 b_{o} \to b_{o1},\;
 b_{1} \to  b_{11},\;
 \hat b_{1} \to \hat b_{11},\;
 b_{4} \to b_{41},\;
\hat b_{4} \to \hat b_{41}.
\end{array}
$$
The explicit form of $K_{s}$, $K_{x_{o}}$, $G_{o}^{m}$ and $g_{1}^{m}$
was given in
\cite{Kuht} by (19).
For $\tilde K_{s}$, $\tilde K_{x_{o}}$, $\tilde G_{o}^{m}$ and $\tilde g_{1}^{m}$ we have

\begin{equation}  
\begin{array}{ll}
\displaystyle
\fl \tilde K_{s}=
\displaystyle
-L_{s}\ln\frac{Q_{1}^{2}}{S-2m^{2}}-
   \\[0.5cm]
\displaystyle
\hspace*{-1.40cm}
-\frac{2}{\sqrt{\lambda_{s}}}\biggr[Li_{2}\biggr(\frac{2m^2}
{S+\sqrt{\lambda_{s}}}\biggl)
-\frac{2}{\sqrt{\lambda_{s}}}\biggr[Li_{2}\biggr(\frac{S+\sqrt{\lambda_{s}}}
{2m^2}\biggl)+\pi^{2}\biggl]
\end{array}
\end{equation}

\begin{equation}  
%\hspace*{0.7cm}
\begin{array}{ll}
\displaystyle
\fl \tilde K_{x_{o}}=
\displaystyle
L_{m}\ln\frac{Q^{2}_{1}Q^{2}}{\lambda_{m}}+
   \\[0.5cm]
\displaystyle
\hspace*{-1.40cm}
+\frac{2}{\sqrt{\lambda_{m}}}\biggr[Li_{2}\biggr(-\frac{4m^{2}Q^{2}}
{(Q^{2}+\sqrt{\lambda_{m}})^{2}}\biggl)
-Li_{2}\biggr(\frac{(Q^{2}+\sqrt{\lambda_{m}})^{2}}
{4m^{2}Q^{2}}\biggl)\biggl]
\end{array}
\end{equation}

\begin{equation}  
%\hspace*{1.7cm}
\begin{array}{ll}
\displaystyle
\fl \tilde G_{o}^{m}=
\displaystyle
-\frac{1}{4}\biggr(L_{mex}\ln\frac{\vert
Q_{1}^{2}\vert}{m^{2}}+
   \\[0.5cm]
\displaystyle
\hspace*{-1.40cm}
+\frac{1}{\sqrt{\lambda_{mex}}}\biggr[Li_{2}\biggr(-\frac{Q_{1}^{2}+\sqrt{\lambda_{mex}}}
{2m^{2}}\biggl)
-Li_{2}\biggr(\frac{2Q_{1}^{2}}
{Q_{1}^{2}+\sqrt{\lambda_{mex}}}\biggl)+\pi^{2}\biggl]\biggl)
\end{array}
\end{equation}

\begin{equation}  
%\hspace*{-3.5cm}
\fl \tilde g_{1}^{m}=
\frac{1}{Q_{1}^{2}+4m^{2}}\biggr(\frac{1}{2}\ln\frac{\vert
Q_{1}^{2}\vert}{m^{2}}
+Q_{1}^{2}\tilde G_{0}^{m}\biggl).
\end{equation}  

The coefficients that appear in (\ref{twodel1})-(73) are:

\hspace*{2cm}
$\Delta=\frac{1}{2}Q^{2}_{1}d_{s}$\;\;\;\;\qquad
$\tilde \Delta=\frac{1}{2}Q^{2}d_{s}$
\begin{eqnarray}
\fl a_{1}=
-\frac{1}{\Delta}\biggr[(K_{x_{o}}+8G_{o}^{m})(X_{o}+2m^{2})\biggl]\qquad&
\hat a_{1}= \frac{1}{\Delta}\biggr[(K_{s}+8G_{o}^{m})(S-2m^{2})\biggl]
 \nonumber \\\displaystyle
\fl a_{11}= -\frac{1}{\tilde \Delta}\biggr[(\tilde
K_{x_{o}}+8\tilde G_{o}^{m})Q^{2}\biggl]&
\hat a_{11}= \frac{1}{\tilde \Delta}\biggr[(\tilde
K_{s}+8\tilde G_{o}^{m})(S-2m^{2}_{e})\biggl]
 \nonumber \\\displaystyle
\fl b_{0}= -\frac{1}{16}\biggr[K_{x_{o}}+8Q^{2}a_{1}\biggl]&
\hat b_{0}= -\frac{1}{16}\biggr[K_{s}+8Q^{2}\hat a_{1}\biggl]
  \\\displaystyle
\fl b_{01}= -\frac{1}{16}\biggr[\tilde K_{x_{o}}+8Q_{1}^{2}a_{11}\biggl]&
\hat b_{01}= -\frac{1}{16}\biggr[\tilde K_{s}+8Q_{1}^{2}\hat a_{11}\biggl]
 \nonumber \\\displaystyle
\fl R_{1}= -\frac{1}{4}L_{x_{o}}-2g_{1}^{m}+Q^{2}a_{1}&
\hat R_{1}= -\frac{1}{4}L_{s}-2g_{1}^{m}+Q^{2}\hat a_{1}
 \nonumber \\\displaystyle
\fl R_{11}= -\frac{1}{4}L_{m}-2\tilde g_{1}^{m}+Q_{1}^{2}a_{11}&
\hat R_{11}= -\frac{1}{4}L_{s}-2\tilde g_{1}^{m}+Q_{1}^{2}\hat a_{11}
 \nonumber 
\end{eqnarray}

\begin{eqnarray}
\hspace*{2cm}
\fl b_{1}=
\frac{1}{\Delta}\biggr[Q^{2}_{1}R_{1}+(g_{1}^{m}-2b_{0})(Q^{2}+4m^{2})\biggl]&
 \nonumber \\\displaystyle
\hspace*{2cm}
\fl \hat  b_{1}=
\frac{1}{\Delta}\biggr[-d_{s}\hat R_{1}+(g_{1}^{m}-2\hat b_{0})(Q^{2}+4m^{2})\biggl]&
 \nonumber \\\displaystyle
\hspace*{2cm}
\fl b_{11}=
\frac{1}{\tilde
\Delta}\biggr[Q^{2}R_{11}+(\tilde g_{1}^{m}-2
b_{01})(Q_{1}^{2}+4m^{2})\biggl]&
 \nonumber \\\displaystyle
\hspace*{2cm}
\fl \hat  b_{11}=
\frac{1}{\tilde \Delta}\biggr[-d_{s}\hat R_{11}+(\tilde g_{1}^{m}-2\hat
b_{01})(Q_{1}^{2}+4m^{2})\biggl]&
  \\\displaystyle
\hspace*{2cm}
\fl b_{4}=
\frac{1}{\Delta}\biggr[Q^{2}_{1}R_{1}+(g_{1}^{m}-2b_{0})(Q^{2}+2X_{o})\biggl]&
 \nonumber \\\displaystyle
\hspace*{2cm}
\fl \hat  b_{4}=
\frac{1}{\Delta}\biggr[-d_{s}\hat R_{1}+(g_{1}^{m}-2\hat b_{0})(2S-Q^{2})\biggl]&
 \nonumber \\\displaystyle
\hspace*{2cm}
\fl b_{41}=
\frac{1}{\tilde
\Delta}\biggr[Q^{2}R_{11}+(\tilde g_{1}^{m}-2
b_{01})(Q_{1}^{2}+2Q_{m}^{2})\biggl]&
 \nonumber \\\displaystyle
\hspace*{2cm}
\fl \hat  b_{41}=
\frac{1}{\tilde \Delta}\biggr[-d_{s}\hat R_{11}+(\tilde g_{1}^{m}-2\hat
b_{01})(2S-Q_{1}^{2})\biggl].&
\nonumber
\end{eqnarray}

\section*{Appendix E}
Here we give schematically the formulae of the cross section
$\sigma^{F}_{R}$ (see (\ref{srf})). This contribution was found by means
of
the
system of analytical calculations REDUCE \cite{Red}. The calculations
were
carried out using the scheme given in \cite{Prep2} (see also
\cite{Kuht}). We have       
\begin{equation}
\sigma^{F}_{R}=\frac{\alpha^{3}S}{\lambda_{s}}\biggr[
\sum_{k=l,e,i}\sigma^{k,u}+\sum_{B}\sum_{T}P_{B}P_{T}\sum_{k=l,e,i}\sigma^{k,p}_{BT}\biggl]
\end {equation}
where
\begin{eqnarray}
\fl \sigma^{l,u}=\frac{1}{Q^{4}}S_{1}^{l,u}+S_{2}^{l,u}+\frac{1}{Q^{2}}S_{3}^{l,u}
\nonumber \\
\fl \sigma^{e,u}=\frac{1}{t_{m}^{2}}S_{1}^{e,u}+S_{2}^{e,u}
+\frac{1}{t_{m}}S_{3}^{e,u}
\nonumber \\
\fl \sigma^{i,u}=\frac{1}{t_{m}}S_{1}^{i,u}+\frac{1}{t_{m}Q^{2}}S_{2}^{i,u}
+S_{3}^{i,u}+\frac{1}{Q^{2}}S_{4}^{i,u}
\\
\fl
\sigma^{l,p}_{BT}=\frac{1}{Q^{4}}S_{1BT}^{l,p}+S_{2BT}^{l,p}+\frac{1}{Q^{2}}S_{3BT}^{l,p}
\nonumber \\
\fl
\sigma^{e,p}_{BT}=\frac{1}{t_{m}^{2}}S_{1BT}^{e,p}+S_{2BT}^{e,p}
+\frac{1}{t_{m}}S_{3BT}^{e,p}
\nonumber \\
\fl
\sigma^{i,p}_{BT}=\frac{1}{t_{m}}S_{1BT}^{i,p}+\frac{1}{t_{m}Q^{2}}S_{2BT}^{i,p}
+S_{3BT}^{i,p}+\frac{1}{Q^{2}}S_{4BT}^{i,p}
\nonumber
\end {eqnarray}
where $t_{m}=X-2m^{2}$.
\noindent The results of calculation of $S^{k,u}_{f}$ ($f$=1, 2, 3, 4) and
$S^{k,p}_{fBT}$ ($f$=1,
2, 3, 4)
were obtained as a set of output REDUCE files 
. Because the obtained exact result is very cumbersome 
we do not present him here.
The whole set of REDUCE files, necessary for calculations, can be found at
the http://www.hep.by/mollerad.htm

\end{document}